\documentclass{aa}  

\usepackage{graphicx}
\usepackage{txfonts}
\usepackage[bookmarks=false,pdftex=false,bookmarksopen=false,pdfborder={0,0,0},colorlinks=true,citecolor=black,linkcolor=black]{hyperref}
\begin{document}

   \title{[Mg/Fe] ratios in the solar neighbourhood: stellar yields and chemical evolution scenarios}


   \author{Marco Palla\inst{1,2,3} \and
          Pablo Santos-Peral\inst{4,5} \and
          Alejandra Recio-Blanco\inst{4}
          \and
          Francesca Matteucci\inst{1,2,6}
          }

   \institute{INAF – Osservatorio Astronomico di Trieste, Via G.B. Tiepolo 11, 34131  Trieste, Italy \\
            \email{marco.palla@inaf.it}
        \and Dipartimento di Fisica, Sezione di Astronomia, Università di Trieste, Via G.B. Tiepolo 11, 34131 Trieste, Italy
         \and 
             Sterrenkundig Observatorium, Ghent University, Krijgslaan 281 - S9, 9000 Gent, Belgium
        \and
            Université Côte d’Azur, Observatoire de la Côte d’Azur, CNRS, Laboratoire Lagrange, Bd de l’Observatoire, CS 34229, 06304 Nice cedex 4, France
        \and
            Departamento de F\'{i}sica de la Tierra y Astrof\'{i}sica, Universidad Complutense de Madrid, 28040 Madrid, Spain
        \and
            INFN – Sezione di Trieste, Via Valerio 2, 34100 Trieste, Italy\\     
             }

   \date{Received xxx; accepted xxx}

 
  \abstract
   {The [Mg/Fe] abundance ratios are a fundamental fossil signature to trace the chemical evolution of the disc and dissecting it into low-$\alpha$ and high-$\alpha$ populations. Despite of the huge observational and theoretical effort, discrepancies between models and data are still present and several explanations have been put forward to explain the [$\alpha$/Fe] bimodality. }
  {In this work, we take advantage of a new AMBRE:HARPS dataset, which provides new and more precise [Mg/Fe] estimations, as well as reliable stellar ages for a subsample of stars, to study the [$\alpha$/Fe] bimodality and the evolution of the solar neighbourhood.}
  {The above data are compared with detailed chemical evolution models for the Milky Way, exploring the most used prescriptions for stellar yields as well as different formation scenarios for the Galactic disc, i.e. the delayed two-infall and the parallel model, also including prescriptions for stellar radial migration.}
  {We see that most of the stellar yield prescriptions struggle to reproduce the observed trend of the data and that semi-empirical yields are still the best to describe the [Mg/Fe] evolution in the thick and thin discs. In particular, most of the yields still predict a steeper decrease of the [Mg/Fe] ratio at high metallicity than what is shown by the data.The bulk of the data are well reproduced by the parallel and two-infall scenarios, but both scenarios have problems in explaining the most metal-rich and metal-poor tails of the low-$\alpha$ data.  These tails can be explained in light of radial migration from inner and outer disc regions, respectively.}
  { Despite of the evidence of stellar migration, it is difficult to estimate the actual contribution of stars from other parts of the disc to the solar vicinity in the data we adopt. However, the comparison between data and models suggests that peculiar histories of star formation, such as that of the two-infall model, are still needed to reproduce the observed distribution of stars.}


   \maketitle
%

\section{Introduction}

Our current understanding of the formation and evolution of our Milky Way (MW) disc is mostly based on the study and interpretation of signatures imprinted in stellar populations. In particular, the stellar atmospheres of non-evolved stars inform us about metals in the interstellar medium (ISM) at the time of their formation (\citealt{Freeman02}), providing for decades the main source of information for the formation and evolution of our own Galaxy. 

The [$\alpha$/Fe] versus metallicity [M/H] (or [Fe/H]) diagram provides valuable clues to the MW disc evolution. The analysis of large survey data, such as APOGEE (e.g. \citealt{Hayden15,Ahumada19}), GALAH (e.g. \citealt{Buder19}) Gaia-ESO (e.g. \citealt{Recio14,Rojas16}), and AMBRE project (e.g. \citealt{Mikolaitis17}) suggests the existence of a clear distinction between two sequences of stars in the [$\alpha$/Fe] vs. [Fe/H] space, i.e. the so-called high-$\alpha$ sequence (often associated to the thick disc) and low-$\alpha$ sequence (associated to the thin disc\footnote{it is worth reminding however that the chemical and geometrical separations give rise to rather different groups (\citealt{Minchev19}, see also \citealt{Caffau21,Romano21}). Therefore, from now on we will refer only to the high-$\alpha$ and low-$\alpha$ sequences.}). 
The significance of this division is even enhanced by the arrival of new dimensions in the data space, such as accurate stellar ages (e.g. from asteroseismology, \citealt{Pinsonneault14,Pinsonneault18,Miglio21}) as well as kinematics and dynamical properties provided by the {\it Gaia} mission (\citealt{Gaia16,Gaia18,Gaia21}).\\

However, the origin and shape of this bimodality is still a matter of debate. 
Several theoretical models suggested that the bimodality may be strictly connected to a delayed gas accretion episode. As for example, by revising the classical two-infall chemical evolution model by \citet{Chiappini97}, \citet{Spitoni19,Spitoni20} showed that a significant delay (4-5 Gyr) between two consecutive episodes of gas accretion is needed to explain the dichotomy in a local stellar sample from APOKASC (\citealt{Silva18}). Similar conclusions were found in \citet{Palla20} and \citet{Spitoni21}, where this scenario was extended to the entire MW disc by comparison with the APOGEE data. 
A late second gas accretion after a prolonged period with a quenched star formation rate (SFR) has also been suggested by the dynamical models of \citet{Noguchi18} and in cosmological simulations (e.g. \citealt{Grand18,Mackereth18}). In this framework, \citet{Buck20} explained the dichotomy as the consequence of a gas-rich merger occurring at a certain epoch in the evolution of the Galaxy (as first proposed in a cosmological model by \citealt{Calura09}).

On the other hand, \citet{Grisoni17} tested the possibility of a picture where high-$\alpha$ and low-$\alpha$ discs are described by two separate and coeval evolutionary sequences, i.e. the parallel scenario. This scenario was found to be particularly viable to explain the presence of a high-$\alpha$ metal rich (H$\alpha$MR) stellar population seen in AMBRE (e.g. \citealt{Mikolaitis14}) as well as in LAMOST (\citealt{Sun20}) data. In the framework of hydrodynamical simulations, the two distinct evolutionary sequences are explained in terms of disc fragmentation: the high-$\alpha$ sequence is attributed to clumps of star formation with high star formation efficiency, whereas the low-$\alpha$ one originates from a more distributed and less efficient star formation episode, as shown in \citet{Clarke19}.

The overall picture on MW disc evolution is further complicated by the existence of stellar radial migration, which is well theoretically established (e.g. \citealt{Sellwood02,Schonrich09,Minchev11,DiMatteo13}) as well as traced in different stellar spectroscopic surveys (e.g. \citealt{Kordopatis15,Hayden17,Minchev18,Feltzing20}). In particular, the presence of stars with very high metallicity ([M/H]$\gtrsim$ 0.1 dex) and circular orbits at solar radius has been interpreted as clear evidence of stellar migration from inner regions of the Galaxy.
However, the real impact of radial migration on the chemical evolution of the Galactic disc is still under debate.
As for example, \citet{Sharma20} presented chemical evolution models\footnote{actually, they presented empirical age-abundance relations for different radii, whose behaviour is not demonstrated to be consistent with the assumed SFR evolution (see also \citealt{Spitoni21,Johnson21}).} where the [$\alpha$/Fe] dichotomy can be reproduced in terms of both the sharp fall of [$\alpha$/Fe] due to the onset of Type Ia SNe and stellar radial migration. In particular, the high-$\alpha$ sequence is seen as a series of stars with different stellar ages, while the low-$\alpha$ one is mostly a series of stars with different birth radii. 
On the contrary, chemical evolution models by \citet{Johnson21} including stellar migration failed to match the observed [$\alpha$/Fe] distribution, indicating that models with smooth star formation histories (SFHs) miss to reproduce the $\alpha$-bimodality (see also \citealt{Vincenzo21}). Also, the results of recent chemo-dynamical simulations has raised doubts about the importance of radial stellar migration in the evolution of abundance ratios in the Galactic disc (e.g. \citealt{Vincenzo20,Khoperskov21}), suggesting the [$\alpha$/Fe]  bimodality is strictly either linked to different star formation regimes or different accretion episodes over the Galaxy lifetime.\\

In this paper, we contribute to this field by comparing detailed chemical evolution models for the solar vicinity with the new local (d$_{sun}$<300 pc) [Mg/Fe] data from \citet{Santos20,Santos21}.\\
In \citet{Santos20} was carried out an optimized normalization procedure for stellar spectra which drive to a decreasing [Mg/Fe] trend at supersolar metallicities, at variance with previous works showing a flat trend for the metal-rich ([M/H]>0) disc (e.g. \citealt{Adibekyan12,Hayden17,Buder19}).\\
The comparison of this new dataset with chemical evolution model is multifold. First of all, we explore how different stellar yield sets behave in reproducing the abundances of \citet{Santos20,Santos21}. In fact, Mg is usually used as $\alpha$-elements tracer (e.g. \citealt{Mikolaitis14,Carrera19}) because of its simple detection in optical spectra, but its modelling in high-mass star ejecta often produce chemical evolution tracks inconsistent with the observations (see, e.g., \citealt{Romano10,Prantzos18}). Furthermore, we compare the data with different scenarios of chemical evolution proposed for the solar neighbourhood, i.e. the parallel model (\citealt{Grisoni17}) and the delayed two-infall (\citealt{Spitoni19,Palla20}), in order to highlight their strengths and weaknesses. In light of these two scenarios, we also consider the possible impact of stellar migration on this local sample of stars. 
To this aim, we implement literature prescriptions for radial migration (from \citealt{Spitoni15,Frankel18}) within our model frameworks. In this way, we test which pathways are likely to explain the stars candidate to be migrated from the inner and outer disc.

The paper is organised as follows. In Section \ref{s:data}, the observational data adopted are presented. In Section \ref{s:models}, we present the main characteristics of model scenarios, as well as the adopted stellar yields and radial migration prescriptions. In Section \ref{s:results}, we show our results and discuss their implications. Finally, in Section \ref{s:conclusions}, we draw our conclusions.

\section{The AMBRE:HARPS observational data sample}
\label{s:data}

\defcitealias{Santos20}{S20}
\defcitealias{Santos21}{S21}

The AMBRE Project (\citealt{deLaverny13}) is a collaboration between the Observatoire de la Côte d’Azur (OCA) and the European Southern Observatory (ESO) to automatically and homogeneously parameterize archived stellar spectra from ESO spectrographs: FEROS, HARPS, and UVES. The stellar atmospheric parameters (T$_{\rm eff}$, log(g), [M/H], [$\alpha$/Fe]) were derived by the multi-linear regression algorithm MATISSE (MATrix Inversion for Spectrum SynthEsis, \citealt{Recio06}), also used in the Gaia RVS (Radial Velocity Spectrometer) analysis pipeline (see \citealt{Recio16}), using the AMBRE grid of synthetic spectra (\citealt{deLaverny12}). 

For the present paper, we only considered the AMBRE:HARPS stellar sample\footnote{The AMBRE analysis of the HARPS spectra comprises the observations collected from October 2003 to October 2010 with the HARPS spectrograph at the 3.6m telescope at the La Silla Paranal Observatory, ESO (Chile).}, which corresponds to stars in the solar neighbourhood observed at high spectral resolution ($R\sim115\,000$, described in \citealt{Pascale14}). 
Concretely, we adopt the abundance data for 1066 stars from the samples of \citet{Santos20} and \citet{Santos21} (hereafter \citetalias{Santos20} and \citetalias{Santos21}). These stars were selected to have T$_{\rm eff}$~>~4700~K, as cooler stars could have larger errors in the parameters (c.f. Fig.~12 in \citealt{Pascale14}). The sample consists in mostly dwarfs (3.3~$\lesssim$~log~g~$\leq$~4.75~cm~s$^{-2}$) cooler than 6200~K, which external uncertainties (estimated by comparison with external catalogues) on T$_{\rm eff}$ , log(g), [M/H], [$\alpha$/Fe] and v$_{\rm rad}$ are 93K, 0.26 cm~s$^{-2}$, 0.08 dex, 0.04 dex and 1~km~s$^{-1}$, respectively. Relative errors from spectra to spectra are much lower. In addition, the stellar [Mg/Fe] abundances were derived following the described methodology in \citetalias{Santos20}, where the spectral normalisation procedure was optimised for the different stellar types and each particular Mg line separately. They present an overall internal error around 0.02 dex and an average external uncertainty of 0.01~dex with respect to four identified Gaia-benchmark stars (18 Sco, HD 22879, Sun, and $\tau$ Cet) from~\citet{Jofre15}. 

It is worth noting that the [M/H] atmospheric parameter is assumed to be equal to the iron abundance value ([Fe/H]) of the star. The notation comes from the atmosphere models. 
The four-dimensional synthetic spectra grid, used as a reference in the analysis includes variations in T$_{\rm eff}$, log g, [M/H] and [$\alpha$/Fe]. In this grid, the individual abundance of all non-$\alpha$ chemical elements (e.g. Fe, Cr, Eu, etc.) is equal to [X$_{non-\alpha}$/H]=[M/H]. On the other hand, the individual abundance of all the $\alpha$-elements [X$_{\alpha}$/H] is scaled following the [$\alpha$/Fe] dimension: [X$_{\alpha}$/H]=[$\alpha$/Fe] + [Fe/H], and taken into account in the estimation of [M/H] through an iterative process (c.f \citealt{Recio06}, \citealt{Recio16}). During the [M/H] estimation, all the non-$\alpha$ metallicity indicators in the observed spectra are considered. These indicators are generally dominated by Fe lines in the optical, as in the case of the AMBRE:HARPS sample. As a consequence, the estimated [M/H] value follows the [Fe/H] abundance with a tight correlation.
This is illustrated in Fig.~9 of \citet{Pascale14}, showing the correlation of the AMBRE:HARPS [M/H] parameter with the estimated [Fe/H] abundance by an independent group (see e.g. \citealt{Sousa08,Sousa11b,Sousa11a,Adibekyan12}) for the same HARPS spectra. Therefore, in the following results and the corresponding interpretation with the chemical evolution models of the [Mg/Fe] versus [M/H] plane, we can assume that [M/H] actually behaves as [Fe/H].

Regarding the [Mg/Fe] abundance estimates, \citetalias{Santos20} showed a significant improvement in precision by carrying out an optimisation of the spectral normalisation procedure, in particular for the metal-rich population ([M/H]~>~0). The followed methodology made it possible to highlight a decreasing trend in the [Mg/Fe] abundance even at supersolar metallicites, partly solving the apparent discrepancies between the observed flat trend in the metal-rich disc (\citealt{Adibekyan12, Hayden15, Hayden17, Buder19}), and the steeper slope predicted by chemical evolution models (\citealt{Chiappini97, Romano10, Spitoni20, Palla20}). 
In this paper, we directly compare these observational data with the new updated chemical evolution models developed by \citet{Palla20} in order to draw more robust conclusions of the Galactic disc evolution scenario.

\subsection*{Orbital properties and  stellar ages} 
\label{s:orbital_parameters}

\begin{figure*}
\centering
\includegraphics[height=90mm, width=\textwidth]{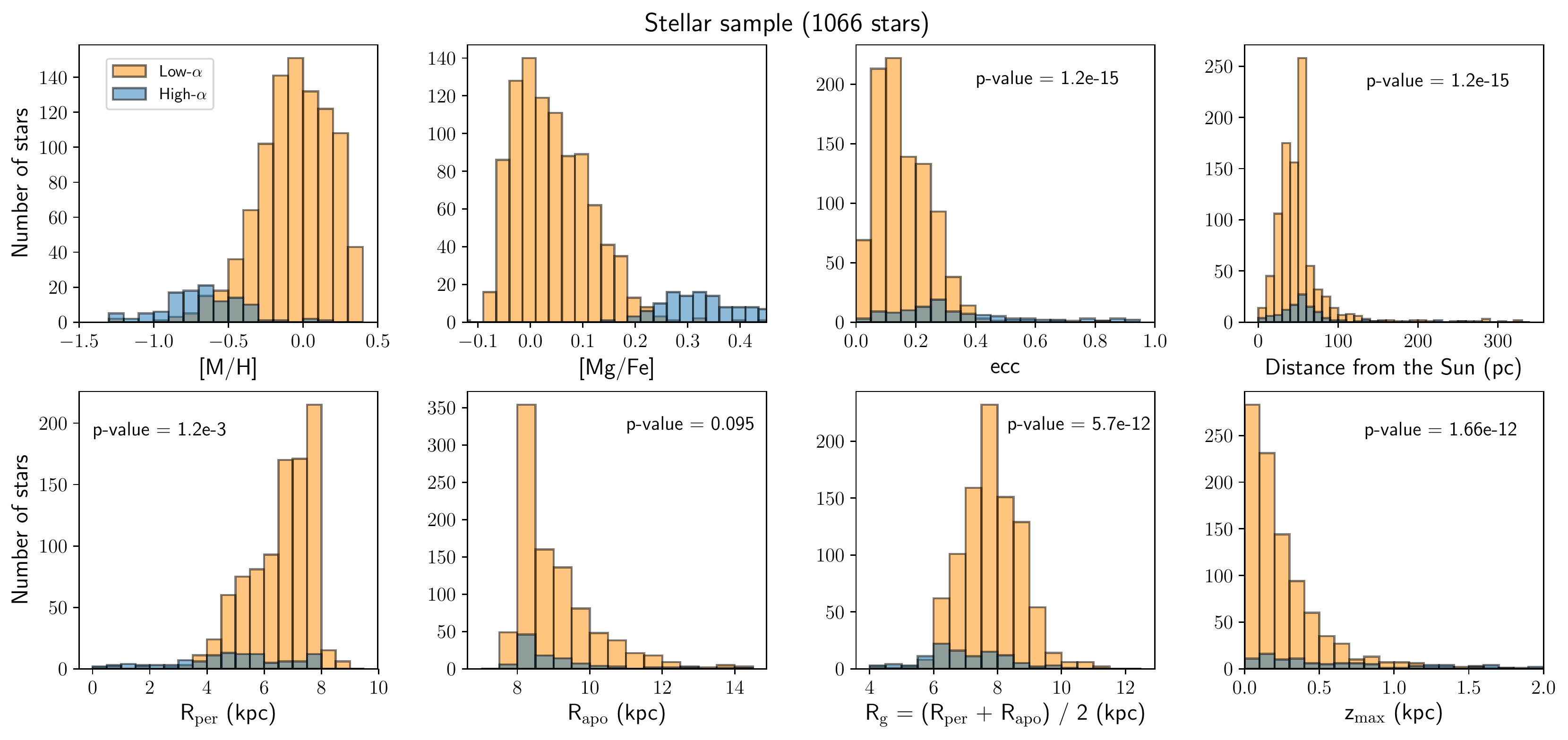}
\caption{Distribution of the main properties ([M/H], [Mg/Fe], eccentricity, distance, R$_{\rm per}$, R$_{\rm apo}$, R$_{\rm g}$, z$_{\rm max}$) of the stellar sample, separated into the chemically defined low-$\alpha$ (orange) and high-$\alpha$ (blue) disc populations. The p-values of the two-sample Kolmogorov-Smirnov tests are shown for the dynamical properties distributions.}
\label{Fig:hist_ALL}
\end{figure*}

As described in \citetalias{Santos21}, a cross-match of the whole AMBRE:HARPS database with the \emph{Gaia} DR2 catalogue (\citealt{Gaia18}) was performed in order to use the accurate astrometric (\citealt{Lindegren18}) and photometric (\citealt{Evans18}) data, along with the estimated distances by \citet{BailerJones18} from \emph{Gaia} DR2 parallaxes, to estimate the orbital parameters and ages of the stars.

Figure~\ref{Fig:hist_ALL} shows the distribution of the main stellar properties of the data sample. These are Galactic disc stars, describing relatively circular prograde orbits (0 < e $\lesssim$ 0.4) close to the Galactic plane (|z$_{\rm max}$|~$\lesssim$~1~kpc), and in the solar vicinity region (d~$\lesssim$~300~pc). We used the guiding centre radius (R$\rm _g$), which was calculated as the average between the pericentre and the apocentre of the orbit of the star: \(\rm R_g = (R_{per} + R_{apo}) / 2 \), as an estimate of the current value of the Galactocentric radius (R$_{\rm GC}$). The sample is well distributed in Galactocentric R$_{\rm g}$ from 4 to 11~kpc.  As the analysed sample is very close to the Sun, the new astrometric information from \emph{Gaia} EDR3 (\citealt{Gaia21}) will not significantly affect our results. 

Moreover, we have reliable age values for a subsample of 366 main sequence turn-off (MSTO) stars, estimated using an isochrone fitting method (\citealt{Kordopatis16}) with PARSEC isochrones (\citealt{Bressan12}). The age distribution for our data subsample ranges from 2.5 to 13.5~Gyr, with an average relative standard deviation of $\sigma~\sim$~20~\% (see complete description in \citetalias{Santos21}).

\subsection*{Disc chemical distinction} 
\label{s:ages}

Here, we adopt a chemical separation of the Galactic disc components, based on the [Mg/Fe] content in a given [M/H] bin.

\begin{figure}
    \centering
    \includegraphics[width=1.\columnwidth]{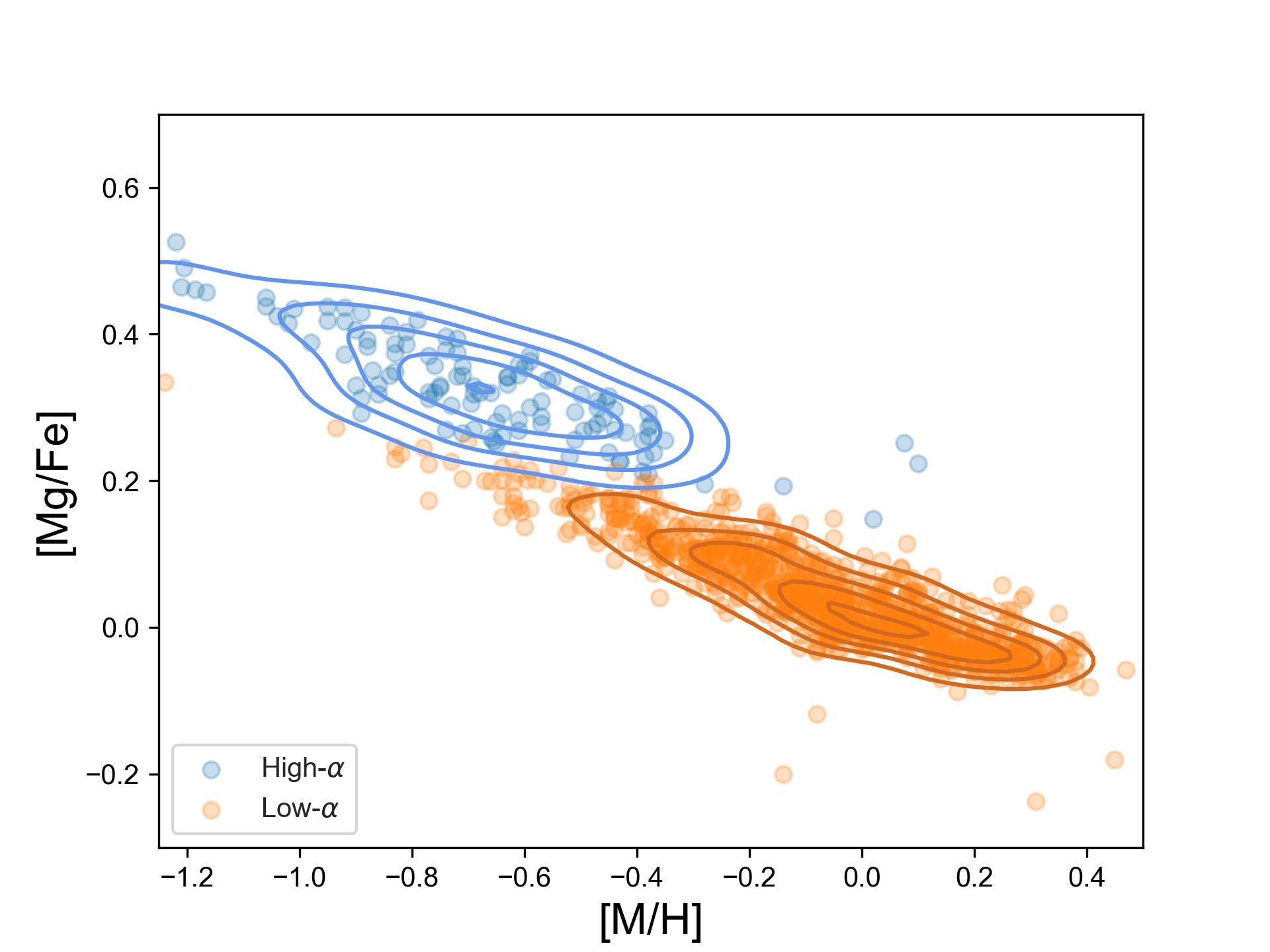}
    \caption{[Mg/Fe] versus [M/H] abundance diagram for the sample presented in \citet{Santos20,Santos21}. The light-blue dots are the observed “high-$\alpha$” stars, whereas the orange dots are the observed “low-$\alpha$” stars. The blue and brown contour lines are the density lines of the observed high-$\alpha$ and low-$\alpha$ stellar distributions, respectively.}
    \label{fig:thick_thin}
\end{figure}

Fig.~\ref{fig:thick_thin} shows the complete sample from \citetalias{Santos20,Santos21}, classified into high-$\alpha$ and low-$\alpha$ sequences. The separation was defined by looking at the gaps in the observed [Mg/Fe] distribution functions in different metallicity bins (of 0.1 dex width) and by performing a linear fit on these divisions (similarly to \citealt{Adibekyan12,Mikolaitis17}). To evaluate the reliability of this division into low and high-$\alpha$ disc populations, we perform a Kolgomorov-Smirnov (KS) test between the divided samples to see if the hypothesis that the two populations are drawn from the same underlying population can be excluded (ilustrated in previous Fig.~\ref{Fig:hist_ALL}). In particular, we look at the dynamical properties (i.e. eccentricity, maximum height above the disc plane, and guiding radius of the orbit) of the stars. The resulting p-values are extremely low (around $<10^{-10}$), confirming that the two populations are actually detached.\\
We also tried to perform other separations. For example, we selected old stars ($t>$ 12 Gyr that would trace the thick disc population, e.g. \citealt{Haywood13}) from the MSTO subsample with stellar ages available, and we draw the lower [Mg/Fe]-bound fit to these stars to define the two sequences. However, as discussed in \citetalias{Santos21}, the result of the KS test on stars with intermediate-high metallicity ([M/H]>-0.3 dex) does not allow to reject the hypothesis of a common origin for the two defined $\alpha$ populations.

In any case, we must say that our chosen separation is rather arbitrary and can assume different shapes/slopes by looking at different data sample: its function is just to give an idea on the high-$\alpha$ and low-$\alpha$ sequences in the abundance diagram plane. Nevertheless, the results given by the KS test confirm the reliability of this division.

\section{Chemical evolution models}
\label{s:models}

In this Section, we present the chemical evolution models adopted in this work. 
To follow the chemical evolution of the solar neighbourhood, we adopt two different scenarios, in order to highlight their advantages and disadvantages in explaining the chemical history of the solar annulus. In the models, this is assumed to be a 2 kpc wide ring centred at ${\rm R}=$ 8 kpc. The models are as follows:
\begin{enumerate}
    \item delayed two-infall model (\citealt{Palla20}). It assumes that the  MW  disc  forms  by  means  of  two sequential  infall  episodes:  the first infall give rise to the high-$\alpha$ sequence, whereas the second (delayed and slower) infall forms the low-$\alpha$ stars. Relative to the "classical" two-infall paradigm (\citealt{Chiappini97,Romano10}), the second infall is delayed by 3.25 Gyr instead of 1 Gyr. The assumption of a delayed second infall allowed us to reproduce abundance diagrams (\citealt{Palla20,Spitoni21}) in the MW disc as well as asteroseismic stellar age trends (\citealt{Spitoni19,Spitoni20}) in the solar neighbourhood. 
    \item parallel model (\citealt{Grisoni17}). It assumes that the two sequences of stars in the chemical space formed out of two separate infall episodes which evolve independently from different gas reservoirs. As for the two-infall model, the infall episode forging high-$\alpha$ stars happens on shorter timescales than the one forming low-$\alpha$ ones. This model was tested to reproduce abundance ratios in the solar neighbourhood (\citealt{Grisoni17}).
\end{enumerate}

For both of the chemical evolution models listed above, the basic equations that describe the chemical evolution of a given element $i$ are:
\begin{equation}
\Dot{G}_i ({\rm R} ,t) = -\psi({\rm R},t) X_i({\rm R},t) + R_i({\rm R},t) + \Dot{G}_{i,inf}({\rm R},t),
    \label{eq:chemical_evo}
\end{equation}
where $G_i (R,t) = X_i ({\rm R},t)\, G({\rm R},t)$ is the fraction of the gas mass in the form of an element $i$ and $G({\rm R},t)$ 
is the fractional gas mass. $X_i({\rm R},t)$ represents the abundance fraction in mass of a given element $i$, with the summation over all elements in the gas mixture being equal to unity.

The first term on the right-hand side of Eq. \eqref{eq:chemical_evo} corresponds to the the rate at which an element $i$ is removed from the ISM due to star formation. The SFR is parametrised according to the Schmidt-Kennicutt law (\citealt{Kennicutt98}):
\begin{equation}
\psi({\rm R},t) = \nu \, \Sigma_{gas}({\rm R},t)^k,
    \label{eq:SFR}
\end{equation}
where $\Sigma_{gas}$ is the surface gas density, $k=1.5$  is  the law index  and $\nu$ is the star formation  efficiency (SFE). This parameter is set to 2 Gyr$^{-1}$ for the first infall/high-$\alpha$ star formation episode and to 1 Gyr$^{-1}$ for the second infall/low-$\alpha$ episode.

$R_{i}({\rm R},t)$ (see \citealt{Palla20b} for the complete expression) takes into account the nucleosynthesis from low-intermediate mass stars (LIMS, $m < 8$ M$_\odot$), core collapse (CC) SNe (Type II and Ib/c, $m > 8$ M$_\odot$) and Type Ia SNe (SN Ia). For these latter, we assume the single-degenerate (SD) scenario and in particular the delay-time-distribution (DTD) by \citet{MatteucciRecchi01}. This choice enables us to obtain abundance patterns that are similar to those obtained with the double degenerate (DD) DTD of \citet{Greggio05} as well as observationally inferred DTDs (e.g. \citealt{Totani08,Maoz17}): for this reason, the SD DTD by \citet{MatteucciRecchi01} can be considered a good compromise to describe the delayed pollution from the entire SN Ia population.\\
$R_i({\rm R},t)$ output is also weighted by the initial mass function (IMF). Here, we adopt the IMF by \citet{Kroupa93}, which is preferred to reproduce the characteristics of the MW disc (\citealt{Romano05}).

The last term in Eq. \eqref{eq:chemical_evo} models the gas infall rate. For the two-infall model, it is computed in this way:
\begin{multline}
    \Dot{G}_{i,inf}({\rm R},t)=A({\rm R})\,X_{i,1inf}({\rm R})\,e^{-\frac{t}{\tau_1}} +\\ +\theta(t-t_{max})\, B({\rm R}) \, X_{i,2inf}({\rm R})\, e^{-\frac{t-t_{max}}{\tau_2({\rm R})}},
    \label{eq:infall_2inf}    
\end{multline}
where $G_{i,inf}({\rm R},t)$ is the infalling material in the form of the element $i$ and $X_{i,Jinf}$ is the infalling gas composition for the $J$-th infall, assumed to be primordial for the solar annulus. $\tau_1$ and $\tau_2$ are the timescales of the two infall  episodes,  while $t_{max}$ indicates  the  time  of  maximum infall,  which  is  also  the  delay  between  the  first  and  the  second infall.  $A$ and $B$ coefficients are set to reproduce the present-day surface mass density of the the chemical discs at a certain radius R. $\theta$ is the Heavyside step function.\\
In the case of the parallel model, since we assume two separate infall episodes, the gas infall law is given instead by these two expressions:
\begin{align}
    \Dot{G}_{i,thick,inf}({\rm R},t)=A({\rm R})\,X_{i,thick,inf}({\rm R})\,e^{-\frac{t}{\tau_1}}\\
    \Dot{G}_{i,thin,inf}({\rm R},t)=B({\rm R})\,X_{i,thin,inf}({\rm R})\,e^{-\frac{t}{\tau_2({\rm R})}}
    \label{eq:infall_parallel}
\end{align}
where the first law refers to the high-$\alpha$ formation episode and the second to low-$\alpha$ stars formation. Here, the different quantities have the same meaning of Eq. \eqref{eq:infall_2inf}. In both the chemical evolution scenarios we set $\tau_1=0.1$ Gyr and and $\tau_2(8$ kpc$)\simeq 7$ Gyr, in agreement with previous studies (e.g. \citealt{Grisoni17,Spitoni19}).\\

During this work, we also run two-infall and one-infall (i.e. low-$\alpha$ star formation episode of the parallel scenario, see Section \ref{ss:inner_outer_results}) models for inner and outer radii to test the impact of radial stellar migration. To this aim, we adopt the prescriptions described in \citet{Palla20} and \citet{Grisoni18}.\\
In particular, we assume that the disc forms inside-out, i.e. the timescale for mass accretion in the Galactic disc increases linearly with radius (\citealt{Chiappini01,Cescutti07}):
\begin{equation}
    \frac{\tau_2({\rm R})}{{\rm Gyr}} \propto \frac{{\rm R}}{{\rm kpc}}.
    \label{eq:inside_out}    
\end{equation}
However, the inside-out mechanism is not sufficient to explain the observed trends for present-day radial gradients (both in abundances and in physical quantities, see \citealt{Palla20}). For this reason, a SFE decreasing with radius is adopted (see also \citealt{Grisoni18}).\\
The total surface mass densities for the low-$\alpha$ and high-$\alpha$ disc are assumed to have exponential profiles. For the two discs, we have that:
\begin{equation}
     \frac{\Sigma({\rm R})}{{\rm M_\odot\, pc^{-2}}}\propto e^{-{\rm R}/{\rm R}_{d}},
    \label{eq:thin_surf}
\end{equation}
where the disc scale length ${\rm R}_d$ is 3.5 kpc for the low-$\alpha$ disc (e.g. \citealt{Spitoni17}) and 2.3 kpc for the high-$\alpha$ disc (\citealt{Pouliasis17}, see also \citealt{Palla20b}). This leads to a $\Sigma_{low}/\Sigma_{high}$ increasing with radius in agreement with recent observational and theoretical works (e.g. \citealt{Anders14,Spitoni21}).\\
For the two-infall model in the inner disc, we also take advantage of the results of \citet{Palla20} and \citet{Spitoni21}, suggesting a metal-enriched gas accretion from the second infall episode, as also predicted by some simulations (e.g. \citealt{Agertz20,Renaud20b,Renaud20}).

We have to mention that in both the models we do not include galactic winds as well as radial gas flows.\\
For the winds, it was found that Galactic fountains rather than wind are more likely to occur in galactic discs. Furthermore, there is the indication (e.g. \citealt{Meioli09,Spitoni09}) that fountains do not alter significantly the chemical evolution of the disc as a whole.\\
Radial gas flows (e.g. \citealt{Spitoni11}) are also neglected in this work, despite of the fact that they were found as viable solutions to explain gradients and abundance ratios (e.g. \citealt{Grisoni18,Palla20}). In fact, most of the works aimed at studying of the solar vicinity assume no flows at all (e.g. \citealt{Romano10,Grisoni17,Spitoni19}). Moreover, tests were performed with radial gas flows, and no substantial differences were found in the general picture.

\subsection{Nucleosynthesis prescriptions}
\label{ss:yield_model}

\defcitealias{Francois04}{F04}
\defcitealias{Koba06}{K06}
\defcitealias{Limongi18}{L18}

The nucleosynthesis prescriptions and the implementation of the stellar yields are fundamental ingredients for chemical evolution models.
Massive stars and SNe Ia play a fundamental role in shaping the [Mg/Fe] vs. [Fe/H] abundance pattern. Mg is mostly produced by massive stars during hydrostatic carbon burning and explosive neon burning (e.g. \citealt{WW95}) whereas SNe Ia are the origin of most of the Fe (e.g. \citealt{Matteucci12}), although CC-SNe produce this element in non-negligible amounts.

In order to see the effects of the nucleosynthesis prescriptions on the evolution scenarios, we decide to test different stellar yields from the literature. For massive stars, we test the following:
\begin{itemize}
    \item \citet{Francois04} (hereafter, \citetalias{Francois04}): empirical modification of solar metallicity \citet{WW95} yields to reproduce the solar Mg abundance. Magnesium yields have been increased in the range 11–20M$_\odot$ and lowered for stars larger than 20M$_\odot$. No modifications were performed for Fe yields computed for solar composition, which do not overestimate Fe abundance as those of \citet{WW95} as functions metallicity (\citetalias{Francois04}).
    \item \citet{Koba06} (hereafter, \citetalias{Koba06}): SN models with different metallicities where $^{56}$Ni output is calibrated on lightcurve and spectral fitting of individual SNe. To satisfy this requirement, hypernovae (HNe, i.e. models with explosion energy ${\rm E}_{SN}>10^{51}$ erg) explosions were also computed for $m>20$ M$_\odot$ in addition to SN models (${\rm E}_{SN}=10^{51}$ erg). Here, we adopt the same set used in \citet{Grisoni17}, where the hypernova fraction ($\epsilon_{HN}$) is set to 1. We highlight that \citetalias{Koba06} table is almost identical for Mg and Fe to those of \citet{Koba11} and \citet{Nomoto13}.
    \item \citet{Limongi18} (hereafter, \citetalias{Limongi18}): SN models with different metallicities where mass loss and stellar rotation are taken into account. In these models, mass cut is chosen in order to allow for the ejection of 0.07 M$_\odot$ of $^{56}$Ni. \citetalias{Limongi18} propose different yield sets for different rotational velocities ($v_{rot}=0$, $150$, $300$ km s$^{-1}$), different minimum mass for failed SNe, i.e. objects fully collapsing to black holes (from 30 M$_\odot$ to 120 M$_\odot$) and with or without fallback and mixing process. At variance with \citetalias{Francois04,Koba06} for which the nucleosynthesis computation does not go beyond 40 M$_\odot$, here yields are computed up to 120 M$_\odot$. Throughout the paper, we will only test the effect different rotational velocities and failed SN masses: the mixing and fallback process do not leave a significant imprint on the analysed abundance patterns.
\end{itemize}

Concerning SNe Ia, we adopt only the yields of the W7 model from \citet{Iwa99}. In fact, \citet{Palla21} shows that negligible differences ($\lesssim$0.1 dex) in [$\alpha$/Fe] ratios are obtained by adopting more recent (and physical) models for different subclasses of SN Ia explosions (e.g. \citealt{Seit13,Leung20}) and that the W7 model can be safely adopted in the analysis of [$\alpha$/Fe] vs. [Fe/H] abundance patterns.

\defcitealias{Spitoni15}{Sp15}
\defcitealias{Frankel18}{F18}

\subsection{Implementing stellar radial migration}
\label{ss:migr_model}

As mentioned in the introduction, we include in the models stellar radial migration prescriptions from the literature to test the origin of candidate "migrators".
We implement migration in the chemical evolution model by adopting two different approaches: one takes the output of hydrodynamical simulations for a MW-like disc, while the other one adopts parametric prescriptions.
In particular, the two implementations are as follows:
\begin{enumerate}
    \item \citet{Spitoni15} (hereafter, \citetalias{Spitoni15}): with this approach, we simply move stars formed at a given radius bin of the model, with a certain age and metal content, to the ring corresponding to the solar annulus (i.e. 8 kpc). To this aim, results of the simulations by \citet{Minchev13} are adopted as a reference. In particular, it is assumed that 10\% and 20\% of the stars born at 4 and 6 kpc, respectively, migrate toward 8 kpc, while 60\% of those born at 8 kpc leave the solar radius. At variance with \citetalias{Spitoni15}, in this work we also consider that 20\% and 10\% of stars born at 10 and 12 kpc migrates towards 8 kpc.\\
    In the scenario of the two-infall model we also assume that stellar migration has effect only for the second gas accretion episode, since our focus for stellar migration is on the low-$\alpha$ sequence.\\
    
    \item \citet{Frankel18} (hereafter, \citetalias{Frankel18}): in this work, migration is seen as a result of a diffusion process treated in a parametrical way. Following \citet{Sanders15} and adapting their parameterization to a Galactocentric radius coordinate, the probability for a star to be currently at a Galactocentric radius ${\rm R}_f$, given that it was born at ${\rm R}_0$ and with age $\tau$, can be written as:
    \begin{equation}
    \ln p({\rm R}_f\, |\, {\rm R}_0, \tau) = \ln(c_3) \, -\frac{({\rm R}_f- {\rm R}_0)^2}{2 \, \sigma_{RM} \, \tau/10\, {\rm Gyr}} ,
    \end{equation}
    where $\sigma_{RM}$ is the radial migration strength and $c_3$ a normalization constant ensuring that stars do not migrate to negative radii (see \citetalias{Frankel18} for details). 
    For $\sigma_{RM}$ we adopt a value of 3.5, very similar to that found in \citetalias{Frankel18} as a result of their Bayesian fitting procedure. We highlight that our expression is marginally different from Eq. (10) in \citetalias{Frankel18}, where the migration normalization is set to 8 Gyr. However, we verify that this change does not bring to significant differences in the results. \\
    As for \citetalias{Spitoni15} prescriptions, we assume stellar migration for the two-infall model only after the onset of the second infall. It is worth noting that also in \citetalias{Frankel18} migration was not considered for an "old disc" (age > 8 Gyr in their case) component. 
    
\end{enumerate}

With both the above prescriptions, we assume that radial migration is only a "passive" tracers of chemical evolution, i.e. the metal enrichment of the ISM in the solar vicinity is practically unaffected by the stars born in other regions. \\
Following \citet{Kordopatis13}, a representative value for the velocity of the migrating stars is 1 km s$^{-1}$, which roughly corresponds to 1 kpc Gyr$^{-1}$. Therefore, only stars with stellar lifetimes $\gtrsim$ 1 Gyr, i.e. corresponding to initial masses below 2 M$_\odot$ (see, e.g. Fig. 3 of \citealt{Romano05}), can travel for distances larger than 1 kpc. Also adopting the parametric approach of \citetalias{Frankel18}, the fraction of stars contributing from other rings than that of the solar vicinity is $\lesssim 0.15$ for times below 1 Gyr. Therefore, we can reasonably assume that most of the metals produced by a stellar generation are ejected rather close to their progenitor birthplace.

\section{Results}
\label{s:results}
We run several models to test the effect of different stellar yields and different scenarios for chemical evolution in comparison with the data from \citetalias{Santos20,Santos21}. In particular, we first concentrate on the outcomes of the different nucleosynthesis prescriptions to see the effect of the different yield sets on the predicted abundance ratios. After that, we move to the comparison of the two-infall and the parallel model scenarios, with their advantages and disadvantages relative to solar neighbourhood data. Finally, we discuss the possible influence of inner and outer disc stars on the solar vicinity.

\subsection{Testing stellar yields}
\label{ss:yield_result}
As mentioned in Section \ref{ss:yield_model}, we test different nucleosynthesis prescriptions for massive stars, whose stellar yields have the largest influence on Mg and Fe evolution.\\

\begin{figure}
    \centering
    \includegraphics[width=1.\columnwidth]{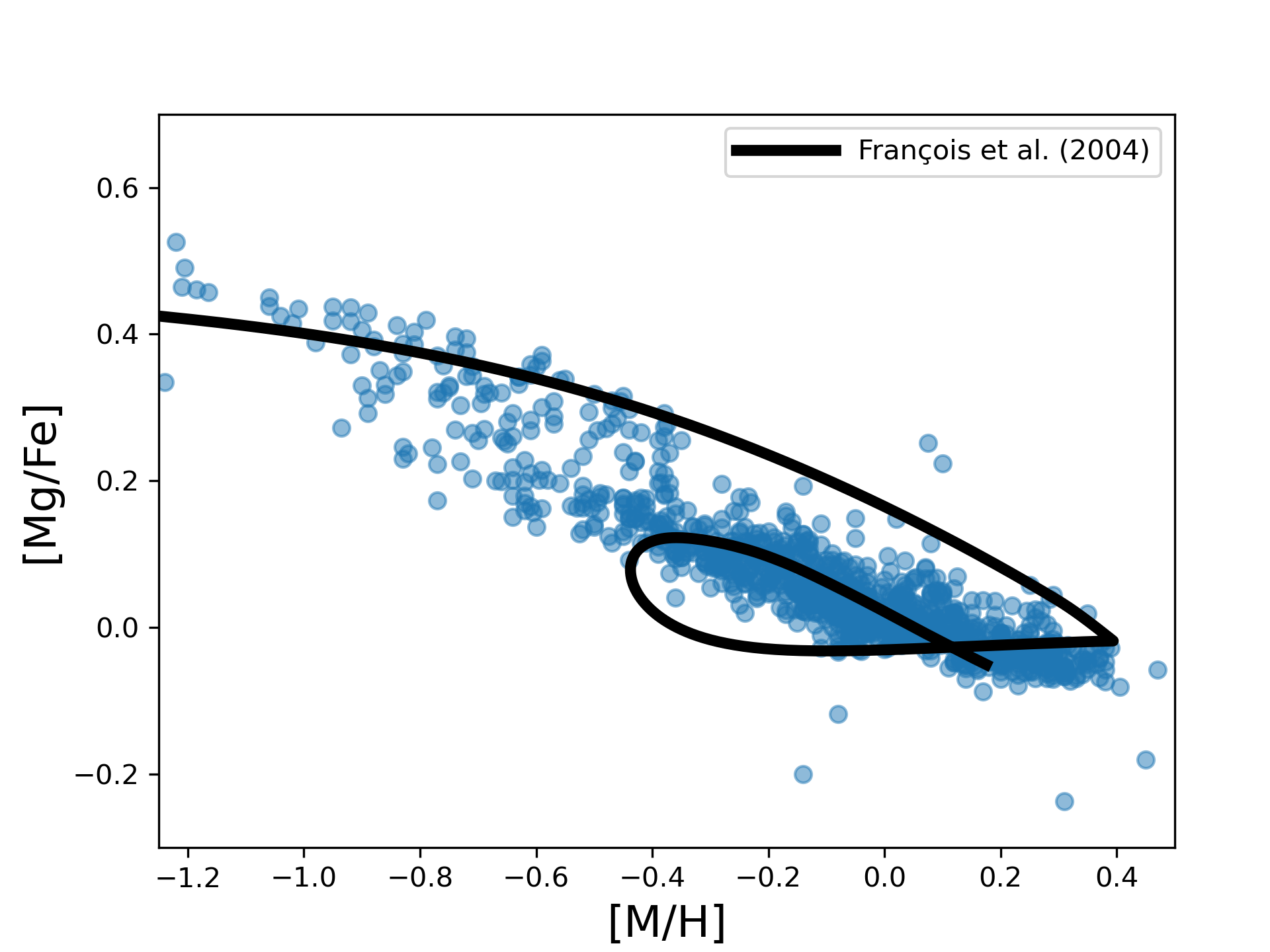}
    \caption{[Mg/Fe] vs. [M/H] abundance diagram for the delayed two-infall model described in Section \ref{s:models} adopting \citet{Francois04} yields for massive stars. Data (blue points) are from \citet{Santos20,Santos21}.}
    \label{fig:Francois}
\end{figure}

In Fig. \ref{fig:Francois}, we plot the predicted [Mg/Fe] vs. [M/H] by the delayed two-infall model adopting the yields of \citetalias{Francois04} and compared with the data from \citetalias{Santos20,Santos21}. 
We see that the model reproduces well the data trend for both the high-$\alpha$ and the low-$\alpha$ sequence. From a more careful inspection, we note however that the model fails to reproduce the region occupied by the low-$\alpha$, low metallicity stars. This is mainly a consequence of the scenario of chemical evolution.

\begin{figure}
    \centering
    \includegraphics[width=1.\columnwidth]{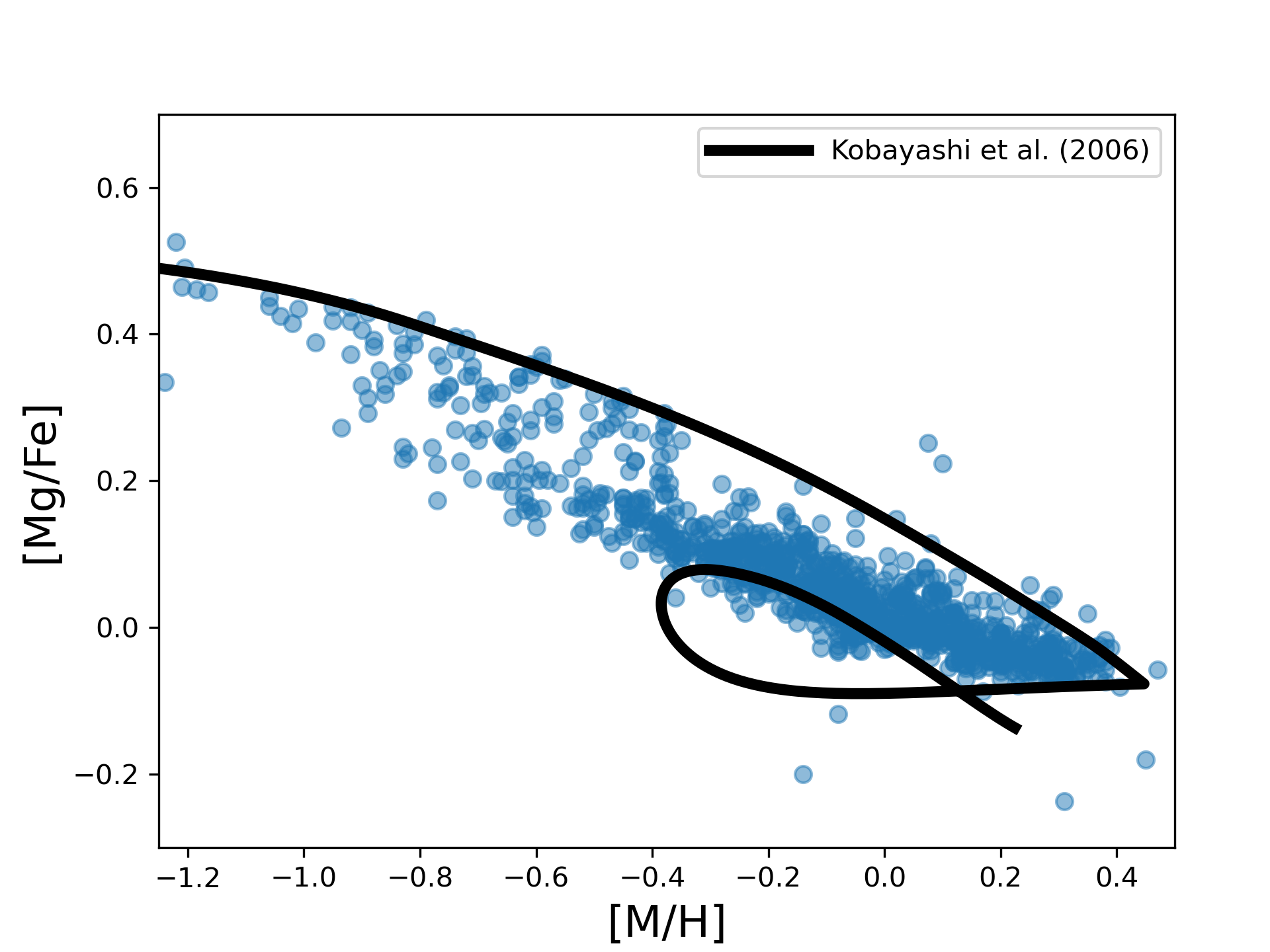}
    \caption{Same of Fig. \ref{fig:Francois}, but for \citet{Koba06} yields.}
    \label{fig:Koba}
\end{figure}

The observational data are also fairly well reproduced by the same chemical evolution model that adopts the CC-SN yields by \citetalias{Koba06}, as shown in Fig. \ref{fig:Koba}. Nevertheless, we see some differences in comparing Fig. \ref{fig:Francois} and \ref{fig:Koba}. The latter Figure shows a model with slightly enhanced [Mg/Fe] ratios at low metallicities, while at supersolar metallicities it emerges a clear underestimation of [Mg/Fe] relative to the trend of the data. This is interesting since \citetalias{Santos20} highlighted for this dataset a decreasing [Mg/Fe] trend with metallicity, at variance with previous estimates (e.g. \citealt{Fuhrmann17,Mikolaitis17}).
Recalling that \citetalias{Francois04} yields are an empirical modification of those of \citet{WW95}, this discrepancy opens the question on whether CC-SN modelling or abundance determination (despite of the optimised derivation procedure) is responsible for this bias.
In fact, it is not possible to attribute the deficiency in the predicted [Mg/Fe] to the SN Ia yields, as often suggested in previous studies (e.g., \citealt{Magrini17,Grisoni18}). This was clearly shown in \citet{Palla21}, where different SN Ia models were found to have negligible effect on the [Mg/Fe] ratio.

\begin{figure}
    \centering
    \includegraphics[width=1.\columnwidth]{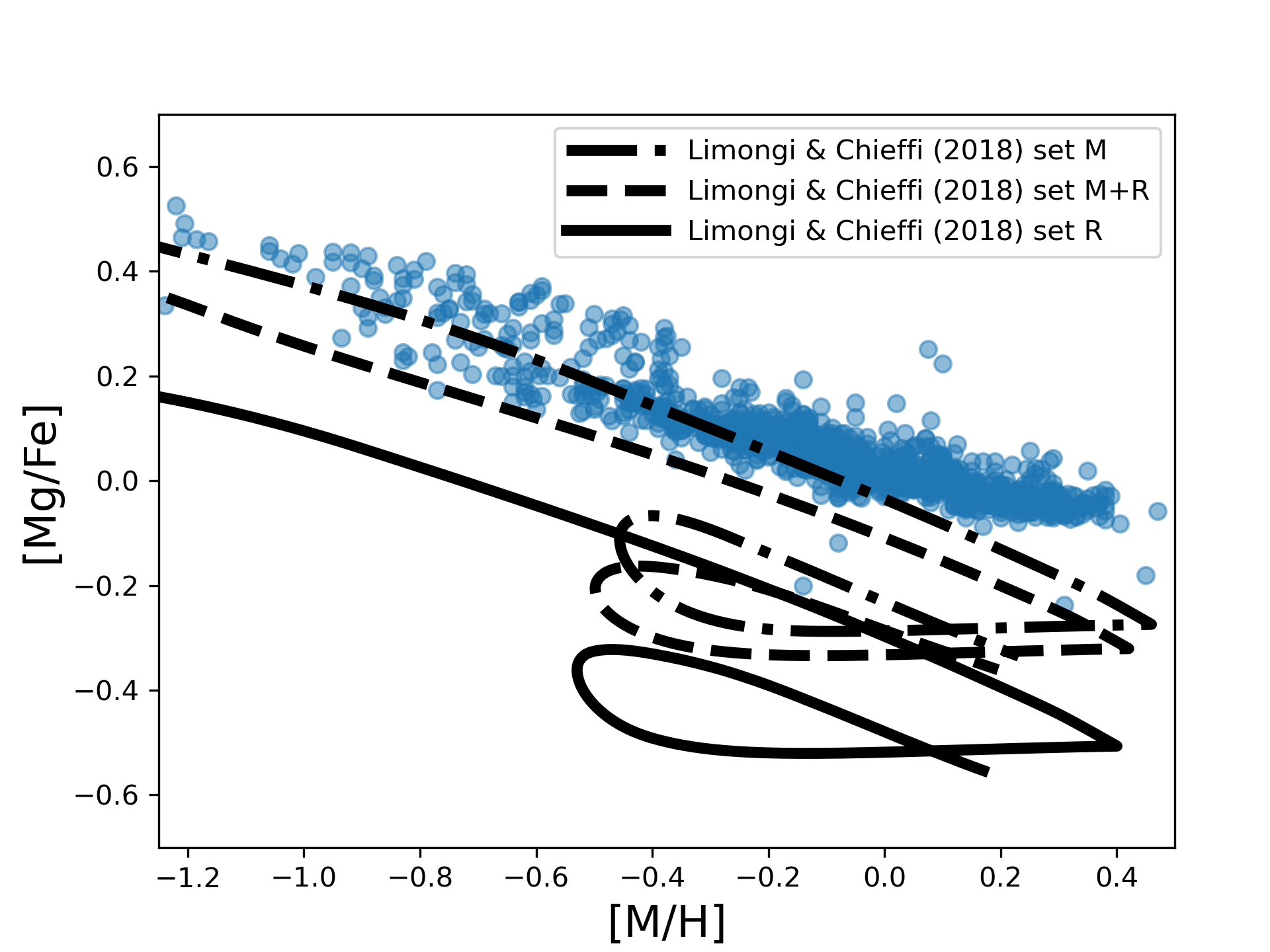}
    \caption{Same of Fig. \ref{fig:Francois}, but for \citet{Limongi18} yields. Solid line: set R with full rotation ($v_{rot}=300$ km s$^{-1}$) up to [Fe/H]=-2, no rotation ($v_{rot}=$0 km s$^{-1}$) afterwards; dashed line: hybrid M+R set (minimum mass for failed SNe >60 M$_\odot$) with full rotation up to [Fe/H]=-2, no rotation afterwards; dash-dotted line: set M with full rotation at all metallicities.}
    \label{fig:Limongi}
\end{figure}

A completely different picture emerges from Fig. \ref{fig:Limongi}. In this Figure we show the delayed two-infall model predictions adopting different sets of stellar yields from \citetalias{Limongi18}.\\
We note that the \citetalias{Limongi18} set R (solid line) severely underestimate the Mg content at all metallicities by at least 0.2 dex. However, this was expected since \citet{Prantzos18} already showed that this nucleosynthesis choice does not agree with the obseved behaviour of [Mg/Fe]. This happens even when adopting yields considering stellar rotation, as those shown in Fig. \ref{fig:Limongi}.
However, \citet{Prantzos18} analysis was limited to the set R of \citetalias{Limongi18}, for which only stars below 30 M$_\odot$ explode as normal CC-SN, whereas more more massive stars end their lives as failed SNe (see Section \ref{ss:yield_model}). In order to probe the differences in relaxing this limit and to be consistent with the other yield sets adopted in this paper (where no failed SNe are assumed), we decide to consider other sets from the compilation of \citetalias{Limongi18}. In particular, we take a hybrid M+R set, where we assume that only stars above 60M$_\odot$ directly collapse to a black hole and the set M, where stars of all masses are considered to die as CC-SNe.
Nevertheless, we note that both the hybrid M+R and the M sets (dashed and dash-dotted line, respectively) are not able to explain the observed [Mg/Fe] vs. [Fe/H] evolutionary trend. Actually, the model adopting the set M decently reproduces the [Mg/Fe] observed at [M/H]$\lesssim$-1 dex, but it lowers to subsolar values for higher metallicities. This happens even adopting yields with full rotation ($v_{rot}=300$ km s$^{-1}$) at solar values: in fact, the dependence on the rotation velocity at this metallicity is very modest (see \citetalias{Limongi18}).\\

Because of the very modest agreement between the data from \citetalias{Santos20,Santos21} and the models adopting the different massive star yields from \citetalias{Limongi18}, from now on we proceed in our analysis adopting the sets from \citetalias{Francois04} and \citetalias{Koba06} only.

\subsection{The different chemical evolution scenarios}
\label{ss:scenario_result}

The stellar yields are not the only factor that influences the predicted chemical evolution for the solar neighbourhood. In recent years, several different scenarios have been proposed to explain the chemical evolution of the MW disc (e.g. \citealt{Nidever14,Haywood16,Grisoni17,Spitoni19,Sharma20}). In this Section, we discuss the results of two different approaches to reproduce the trend observed by \citetalias{Santos20,Santos21}: i) the delayed two-infall model (\citealt{Spitoni19} and later \citealt{Palla20}) and ii) the parallel model (\citealt{Grisoni17}, see Section \ref{s:models} for more details).\\

\begin{figure}
    \centering
    \includegraphics[width=1.\columnwidth]{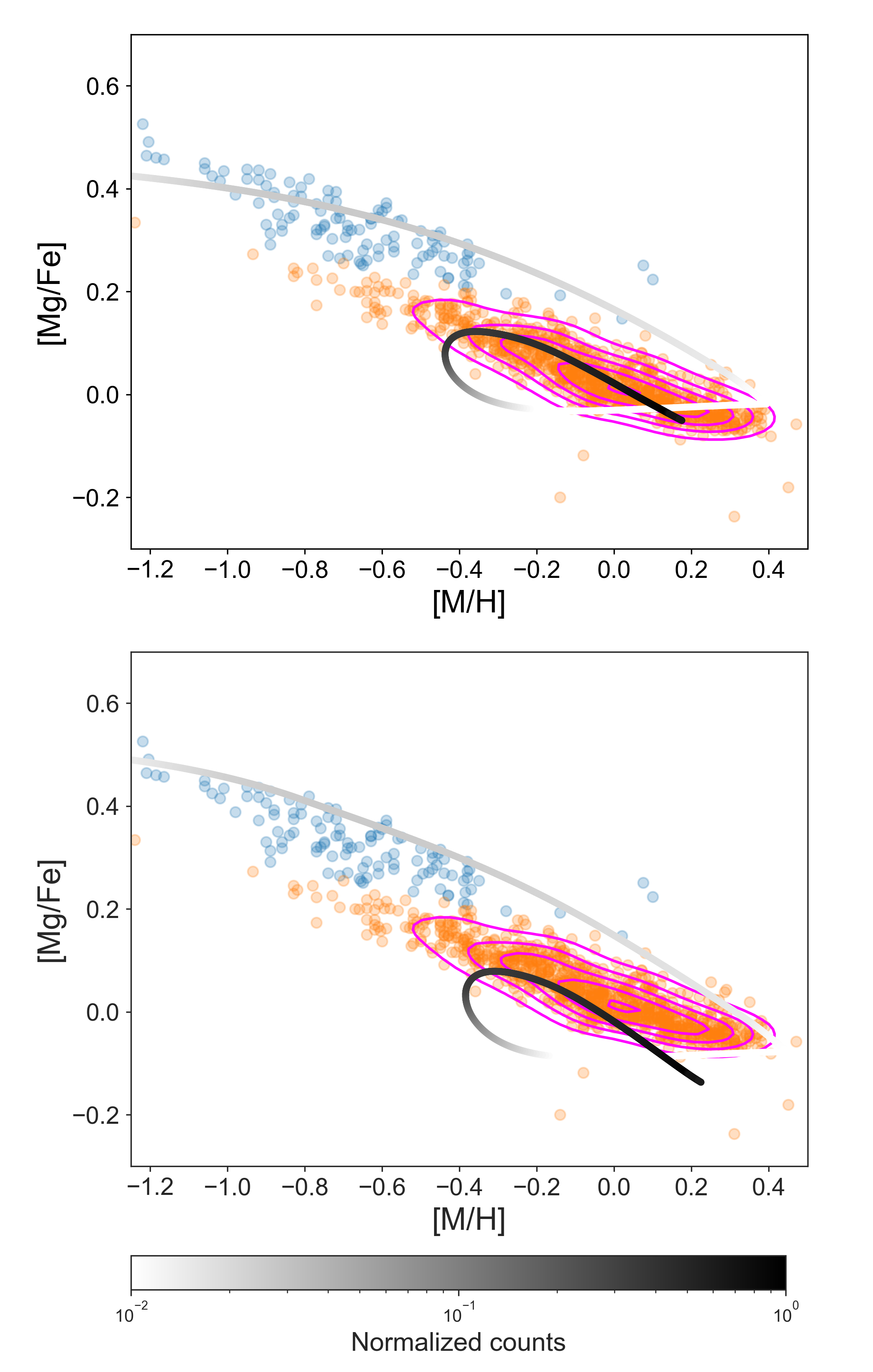}
    \caption{[Mg/Fe] versus [M/H] abundance diagram for the delayed two-infall model scenario of chemical evolution with \citet{Francois04} yields (upper panel) and \citet{Koba06} yields (lower panel). The colour bar indicates the predicted stellar number counts (normalised to the maximum value) at a certain point of the diagram, taking into account the selection function. The light-blue points are the observed “high-$\alpha$” stars, whereas the orange points are the observed “low-$\alpha$” stars. The magenta contour lines enclose the observed density distribution of stars. }
    \label{fig:delayed_twoinfall}
\end{figure}

\begin{figure}
    \centering
    \includegraphics[width=1.\columnwidth]{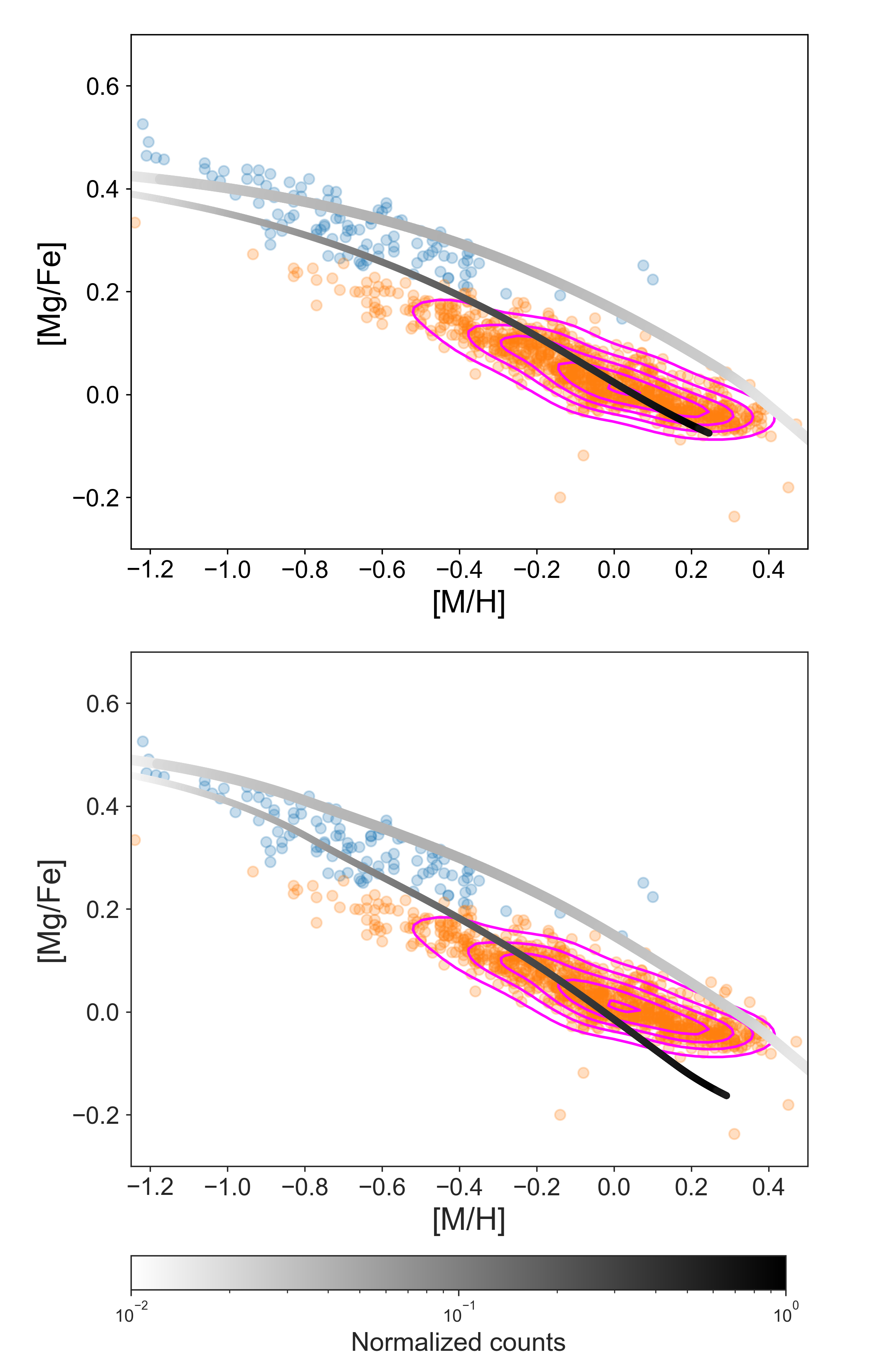}
    \caption{Same of Figure \ref{fig:delayed_twoinfall}, but for the parallel model scenario of chemical evolution. The thin lines represent the chemical track for the low-$\alpha$ star formation episode, whereas the thin lines are the chemical track for the high-$\alpha$ star formation episode. }
    \label{fig:parallel}
\end{figure}

In Fig. \ref{fig:delayed_twoinfall} and \ref{fig:parallel}, we show the predicted chemical evolution by the two scenarios proposed above, adopting the two different sets of stellar yields selected in Section \ref{ss:yield_result}.\\
It is worth noting that to compare the predictions of the models with observations, we take the selection function of the observational data (see Section \ref{s:data}) into account. 
Since this function is age and metal dependent, because the fraction of stars within the selection window changes with the age and metallicity, we include in our models a selection fraction $f(\tau,[M/H])$ computed in this way:
\begin{equation}
    f(\tau,[M/H])=\frac{\int^{M_b(\tau,[M/H])}_{M_a(\tau,[M/H])}\phi(m)\,dm}{ \int^{M_{max}(\tau)}_{M_{min}}\phi(m)\, dm},
\end{equation}
where $\phi(m)$ is the IMF and $M_{a}$ and $M_{b}$ are the initial masses of the stars that are in the $M_J$, log($g$) and $T_{eff}$ ranges for a population of age $\tau$ and metallicity [M/H].
To do this, we adopt PARSEC (release v1.2s) + COLIBRI (release S37) stellar evolutionary tracks (\citealt{Bressan12,Chen14,Chen15,Tang14,Marigo17,Pastorelli19,Pastorelli20}) for selecting the range of masses in the selection window at different ages and metallicities computed by the chemical evolution model.

Looking at Fig. \ref{fig:delayed_twoinfall}, we see that both the \citetalias{Francois04} (upper panel) and \citetalias{Koba06} (lower panel) stellar yields reproduce the trends of the high-$\alpha$ and the low-$\alpha$ sequences as defined in Section \ref{s:data}.
The model also well predicts the bulk of the stellar distribution in the [Mg/Fe] vs. [M/H] plane: especially in the case adopting \citetalias{Francois04} yields, the peak in the stellar counts (see Fig. \ref{fig:delayed_twoinfall} colour bar) coincides with the densest data region of the diagram, as described by the magenta contours. 
However, it can be seen that both models do not explain the low-$\alpha$, low metallicity ([M/H] $\lesssim$ -0.6 dex)  data as well as the high metallicity tail (above [M/H]$\sim$0.2 dex) of the low-$\alpha$ sequence. In fact, even if the models cover the high-metallicity region at the end of the first infall, the number of stars predicted by the simulation at that point is negligible. This is evident by looking at Fig. \ref{fig:delayed_twoinfall} color bar.\\
As noted in Section \ref{ss:yield_result}, the two-infall model with \citetalias{Koba06} yield (Fig. \ref{fig:delayed_twoinfall} lower panel) also underestimates the [Mg/Fe] values at high metallicity. This happens in spite of the results of \citetalias{Santos20}, who obtained a decreasing trend in the [Mg/Fe] data even at supersolar metallicity, differently from previous estimates (e.g. \citealt{Fuhrmann17,Mikolaitis17}).

In Fig. \ref{fig:parallel} we show the results for the parallel model adopting the yields of \citetalias{Francois04} (upper panel) and \citetalias{Koba06} (lower panel).
By looking at the chemical tracks for the high-$\alpha$ star formation episode (thick lines), we note that the this sequence explain well the stellar distribution observed for the high-$\alpha$ stars with both the nucleosynthesis prescriptions adopted. At variance with the delayed two-infall model, we see that the parallel scenario helps to explain the scatter observed for the high-$\alpha$ stars.
Similarly to Fig. \ref{fig:delayed_twoinfall} with the two-infall model, the parallel scenario in Fig. \ref{fig:parallel} has problems in explaining the low metallicity ([M/H] $\lesssim$ -0.6 dex), low-$\alpha$ stars. Moreover, in both upper and lower panels we see that the thick disc model track can explain only a minimal part of the highest metallicity ([M/H] $\gtrsim$ 0.2 dex) stars, due to the very low stellar density predicted by the model (see Fig. \ref{fig:parallel} colour bar). Nonetheless, the parallel scenario is able to explain the bulk of observed stars by \citetalias{Santos20,Santos21} (see magenta contours) adopting either the stellar yields of \citetalias{Francois04} and \citetalias{Koba06}.\\
However, with the latter yield set (Fig. \ref{fig:parallel} lower panel), it is evident that the model for the thin disc underestimates the predicted [Mg/Fe] at supersolar metallicity. The same feature is visible for the two-infall scenario when we adopt the prescriptions of \citetalias{Koba06}. This confirms even more that the high metallicity subsolar Mg obtained with the \citetalias{Koba06} set is a yield-dependent feature that does not depend on the adopted chemical evolution scenario.

\subsubsection*{Considering stellar ages}
\label{sss:ages}

In order to probe the scenarios of chemical evolution on firmer bases, we decide to compare our models with the subsample of MSTO stars (see Section \ref{s:data}) for which accurate stellar ages were derived. The further dimension available for this subsample provide an important test for the two scenarios of chemical evolution, highlighting features that are hidden when looking at the abundance diagrams only.\\

To avoid biased conclusions in the comparison, we select the predicted stars according to the selection function for the MSTO subsample (see \citetalias{Santos21} for the details).\\
Moreover, we take into account the uncertainties in stellar age and abundance determinations by building a synthetic chemical evolution model, i.e. a mock stellar catalog of surviving stars.
Similarly to the procedure adopted in \citet{Spitoni19}, we add at each Galactic time a random error to the ages and the abundances of the stars formed at the Galactic time $t$. The random errors are uniformly distributed in the interval of the average error estimated for a certain abundance and time. In this way, for the synthetic model, we have a "new Age", defined as:
\begin{equation}
    \tau_{new}(t)=\tau(t)+U([-\sigma_{\tau(t)}, +\sigma_{\tau(t)}]),
    \label{eq:Agenew}
\end{equation}
where $\tau$($t$)=(13.7-$t$) Gyr, $U$ is the random function with uniform distribution and $\sigma_{\tau(t)}$ is the average error in the age determination at the time $t$. This value spans a range between $\sim$0.25 to $\sim$2 Gyr. In parallel, we implement the error in the metallicity [M/H] to define a "new metallicity":
\begin{equation}
    [M/H]_{new}(t)=[M/H](t)+U([-\sigma_{[M/H]}, +\sigma_{[M/H]}]),
    \label{eq:MHnew}
\end{equation}
where $\sigma_{[M/H]}$ is independent of age and metallicity and has a value of 0.08 dex.
The same is done for [Mg/Fe], for which we define 
\begin{multline}
    [Mg/Fe]_{new}(t,[Mg/Fe])=[Mg/Fe](t)+\\
    +U([-\sigma_{[Mg/Fe]([Mg/Fe])}, +\sigma_{[Mg/Fe]([Mg/Fe])}]).
    \label{eq:MgFenew}
\end{multline}
In this case, $\sigma_{[Mg/Fe]([Mg/Fe])}$ is the average error in a [Mg/Fe] bin, with typical values of the order of 0.02-0.03 dex (see Section \ref{s:data}).\\

\begin{figure*}
    \centering
    \includegraphics[width=1.\textwidth]{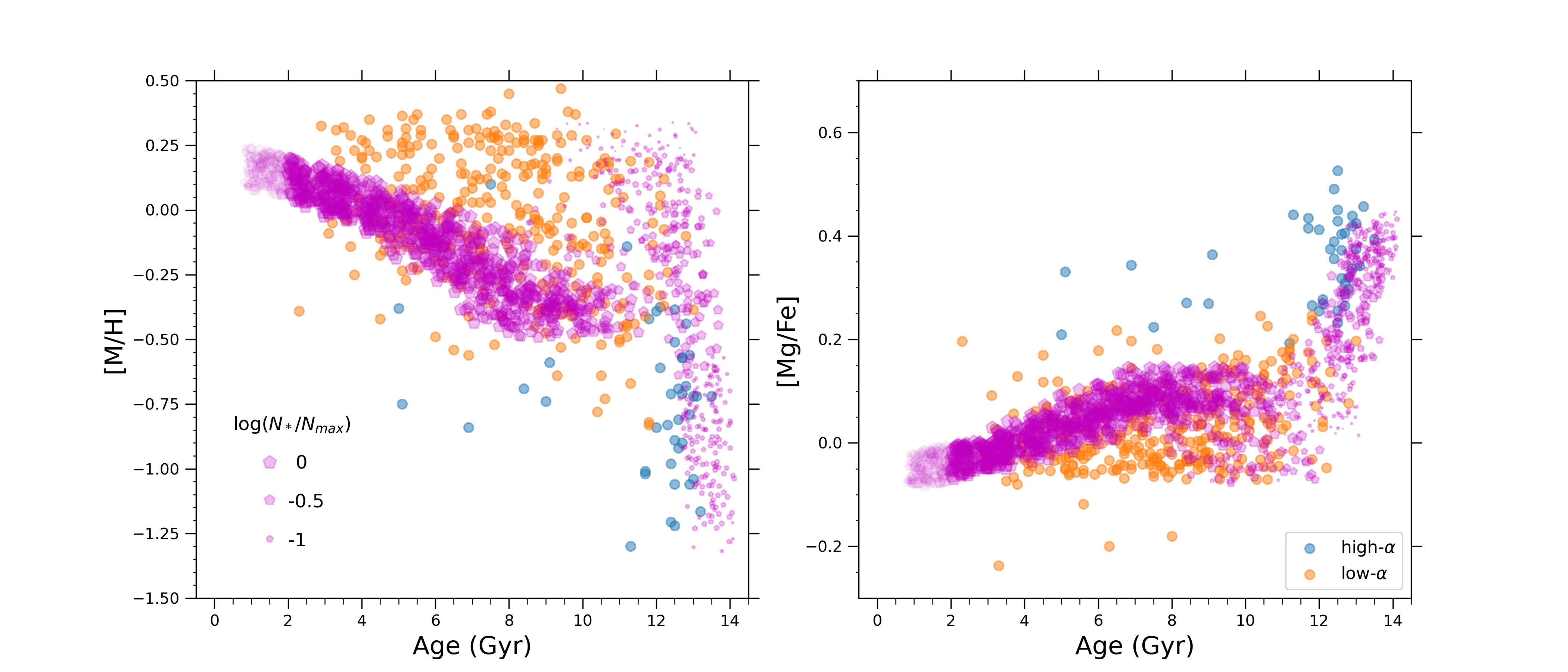}
    \caption{Synthetic model results  for  the  time  evolution  of [M/H] (left panel) and [Mg/Fe] (right panel) distributions produced by the delayed two-infall model in combination with the age and abundance uncertainties of the MSTO subsample (see Eqs. \eqref{eq:Agenew}, \eqref{eq:MHnew} and \eqref{eq:MgFenew}). The stellar yields are from \citet{Francois04}. Purple filled pentagons are the mock data from the synthetic model. The pentagon size indicates  the  local  number  density  of  formed  stars  normalised  to  maximum value, taking into account the selection function. Pentagons with age below 2 Gyr are shaded due to the absence of stellar data in this region. The light-blue points are the observed “high-$\alpha$” stars, whereas the orange points are the observed “low-$\alpha$” stars of the MSTO subsample.}
    \label{fig:age_MH_2inf}
\end{figure*}

In Fig. \ref{fig:age_MH_2inf} the results of the time evolution of [M/H] (left panel) and [Mg/Fe] (right panel) ratios including the errors described in Eqs. \eqref{eq:Agenew}, \eqref{eq:MHnew} and \eqref{eq:MgFenew} are reported for the delayed two-infall model. 
To avoid any confusion in the reader, from now on we plot only the results obtained using \citetalias{Francois04} yields, which result the best in explaining \citetalias{Santos20,Santos21} abundance data. However, we find similar conclusions even when \citetalias{Koba06} prescriptions are adopted.\\
The analysis of Fig. \ref{fig:age_MH_2inf} confirms the problems of the two-infall model already found in Fig. \ref{fig:delayed_twoinfall}. The model fails to reproduce a few of the low-$\alpha$, low-metallicity stars, but mostly does not predict the stars with the highest metallicity. This is evident in Fig. \ref{fig:age_MH_2inf} left panel, where we also note that these high-metallicity stars (hereafter super metal rich, SMR, [M/H]$\gtrsim$0.1 dex)  have a large spread in stellar age, i.e. from 10 Gyr to the most recent ages in the sample ($\sim$ 2 Gyr). We highlight that this spread can be hardly accounted for a scenario of homogeneous chemical evolution in a single disc zone, even accounting for observational errors. Moreover, we must say that large intrinsic scatter for stars of the same age and birthplace is not expected (see, e.g. \citealt{Ness19,Sharma20b}): hence, we can attribute this spread to outward radial migration for a fraction of stars of the MSTO sample.
This idea is supported also by the right panel, where quite large scatter ($\sim$0.2 dex) is seen for [Mg/Fe] for $\tau<$10 Gyr. In this case, the model does not well reproduce the low [Mg/Fe] stars with intermediate ages, that correspond to the high metallicity stars in that age range.\\
In the two panels of Fig. \ref{fig:age_MH_2inf}, we also note that some classified high-$\alpha$ stars (blue points) show very young ages, in contrast to the majority of the objects of this class, which show $\tau>$10 Gyr. These can be accounted for as the so-called "young $\alpha$ rich" (Y$\alpha$R) stars, whose origin is still debated and attributed either to stars migrated from the Galactic bar (e.g. \citealt{Chiappini15}) or more probably evolved blue stragglers (e.g. \citealt{Martig15,Yong16,Jofre16,Izzard18,Zhang21}).

\begin{figure*}
    \centering
    \includegraphics[width=1.\textwidth]{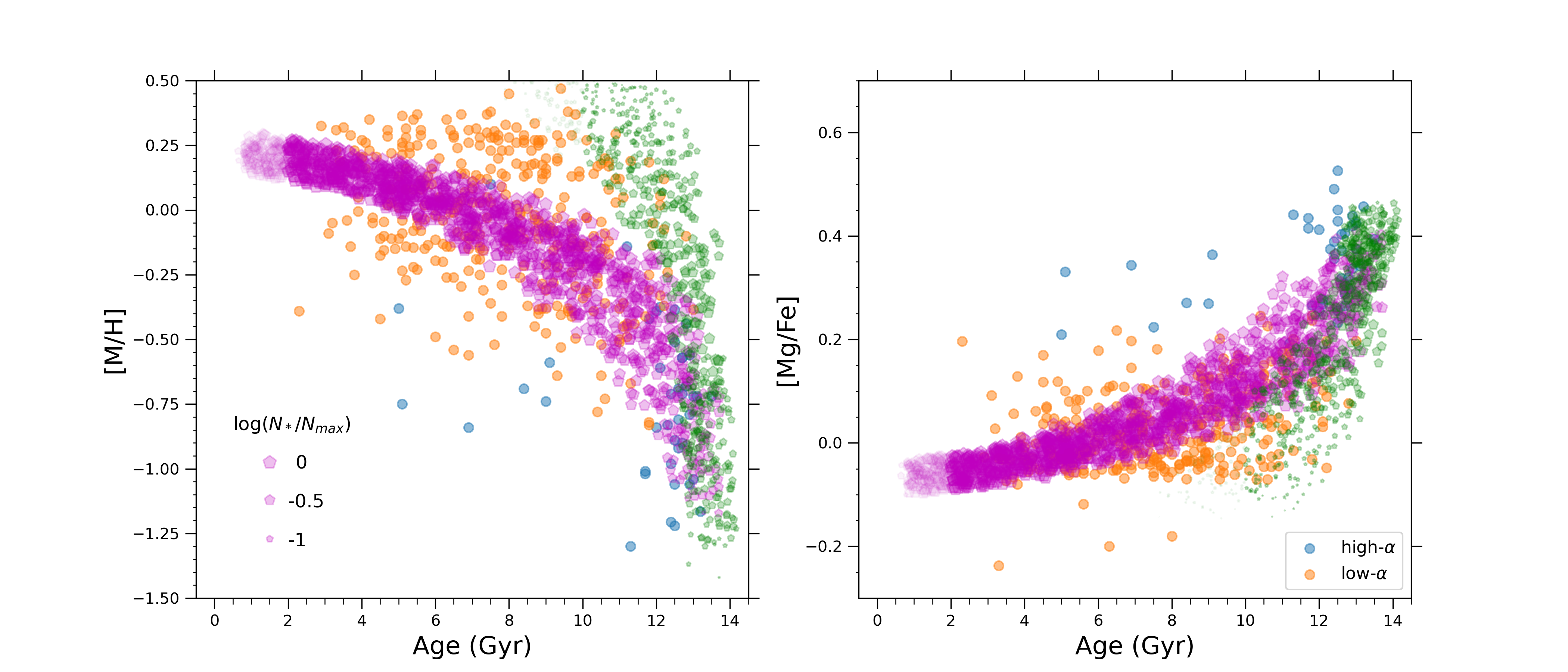}
    \caption{Same of Fig. \ref{fig:age_MH_2inf}, but for the parallel model. The adopted yields are from \citet{Francois04}. Green filled pentagons are the mock data of the high-$\alpha$ sequence of the synthetic model, whereas purple filled pentagons indicate the mock data of the model low-$\alpha$ sequence. In the former case, the pentagons are shaded if the age is below 10 Gyr because the contribution to the total stellar density of lower aged stars is negligible. }
    \label{fig:age_MH_parallel}
\end{figure*}

In Fig. \ref{fig:age_MH_parallel}, we show again the time evolution of [M/H] (left panel) and [Mg/Fe] (right panel) ratios including the errors described in Eqs. \eqref{eq:Agenew}, \eqref{eq:MHnew} and \eqref{eq:MgFenew}, but for the parallel scenario of chemical evolution.\\
In the left panel of Fig. \ref{fig:age_MH_parallel}, we observe that the general scenario is similar to that of the two-infall model. The model fails to reproduce the observed low metallicity, intermediate-young age stars as well as most of SMR stars. However, it is clear that this scenario has more problems in predicting the low metallicity stars relative to the two-infall model in Fig. \ref{fig:age_MH_2inf}. On the contrary, the parallel model has less problems on the side of intermediate-young age, high-metallicity stars. This difference between the parallel and the two-infall scenarios is mostly caused by the higher metallicities reached by the evolutionary track with lower $\alpha$s than that by the second infall episode at similar ages. As can be seen in Fig. \ref{fig:age_MH_parallel}, the high-$\alpha$ track does not help much in reproducing high-metallicity stars. In fact, by looking at the pentagon sizes of Fig. \ref{fig:age_MH_parallel}, the predicted number of stars after the first Gyr of evolution is too low, as already noted in Fig. \ref{fig:parallel}.\\
In the right panel of Fig. \ref{fig:age_MH_parallel}, we see that also in the case of the parallel model the spread in [Mg/Fe] is hardly reproduced by the models. At variance with the two-infall scenario, the "lower $\alpha$" track sequence predicts lower [Mg/Fe]: as for Fig. \ref{fig:age_MH_parallel} left panel, this can be attributed in part to the higher metal enrichment at similar age of this scenario (see \citealt{Matteucci03,Matteucci12}). However, the model still struggle to reproduce the lowest [Mg/Fe] data at intermediate ages.\\

The analysis of the sample of the MSTO stars including stellar age estimates seems to indicate that a fraction of these stars have experienced a radial migration from the place of birth.

Even though the MSTO subsample is not complete, the indications coming from Figs. \ref{fig:age_MH_2inf} and \ref{fig:age_MH_parallel} are quite evident. In particular it seems, from this analysis, that most of the observed SMR (and lowest [Mg/Fe]) stars observed in the solar vicinity should have been formed in faster chemically evolving environment, such as that of the inner disc (see also \citealt{Minchev13,Kordopatis15}).\\
This suggestion is also supported by looking at the stellar specific angular momentum behaviour with metallicity in the sample of \citetalias{Santos20,Santos21}. In fact, we observe that the mean angular momentum for SMR stars is anti-correlates with [M/H]: this is a clear consequence of radial migration from the innermost regions of the disc (see also \citealt{Sharma20}). The mean angular momentum increases with radius, because $J_{\phi}=v_{circ}\, {\rm R}$: therefore, the lower the radius, the lower the angular momentum.

\subsection{Impact of inner and outer disc on the solar neighbourhood}
\label{ss:inner_outer_results}

The difficulties encountered by both the chemical evolution scenarios in predicting low-$\alpha$, low metallicity stars and SMR stars (see Section \ref{ss:scenario_result}) point towards a context where these stars were born at different Galactic locations from the solar vicinity.

In this Section, we test which can be the origin of these stars in the light of the scenarios of chemical evolution described in Section \ref{s:models}. \\
First, we take advantage of the works of \citet{Grisoni18} and \citet{Palla20} (see also \citealt{Spitoni21}), which proposed parameterisations for different Galactocentric radii within the framework of the one-infall model\footnote{in this Section we do not consider the parallel scenario but rather the one-infall model. This model traces the formation of low-$\alpha$ sequence only.} and the delayed two-infall model. Having tested if these chemical pathways are viable to describe the low and high metallicity tails of the low-$\alpha$ sequence, we implement  stellar radial migration prescriptions (\citealt{Spitoni15,Frankel18}, see \ref{ss:migr_model}) to probe the proposed scenarios.\\

\begin{figure}
    \centering
    \includegraphics[width=1.\columnwidth]{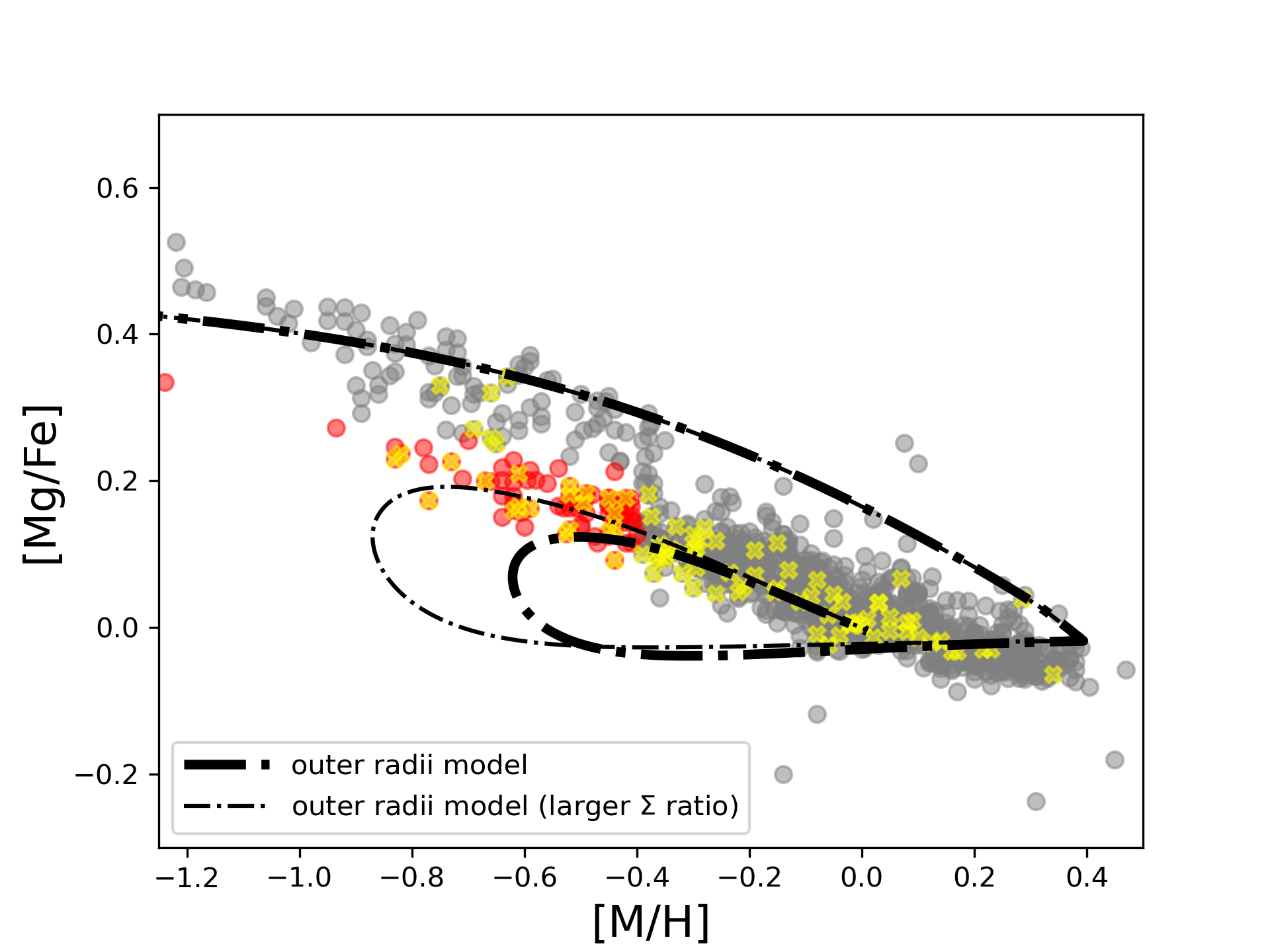}
    \includegraphics[width=1.\columnwidth]{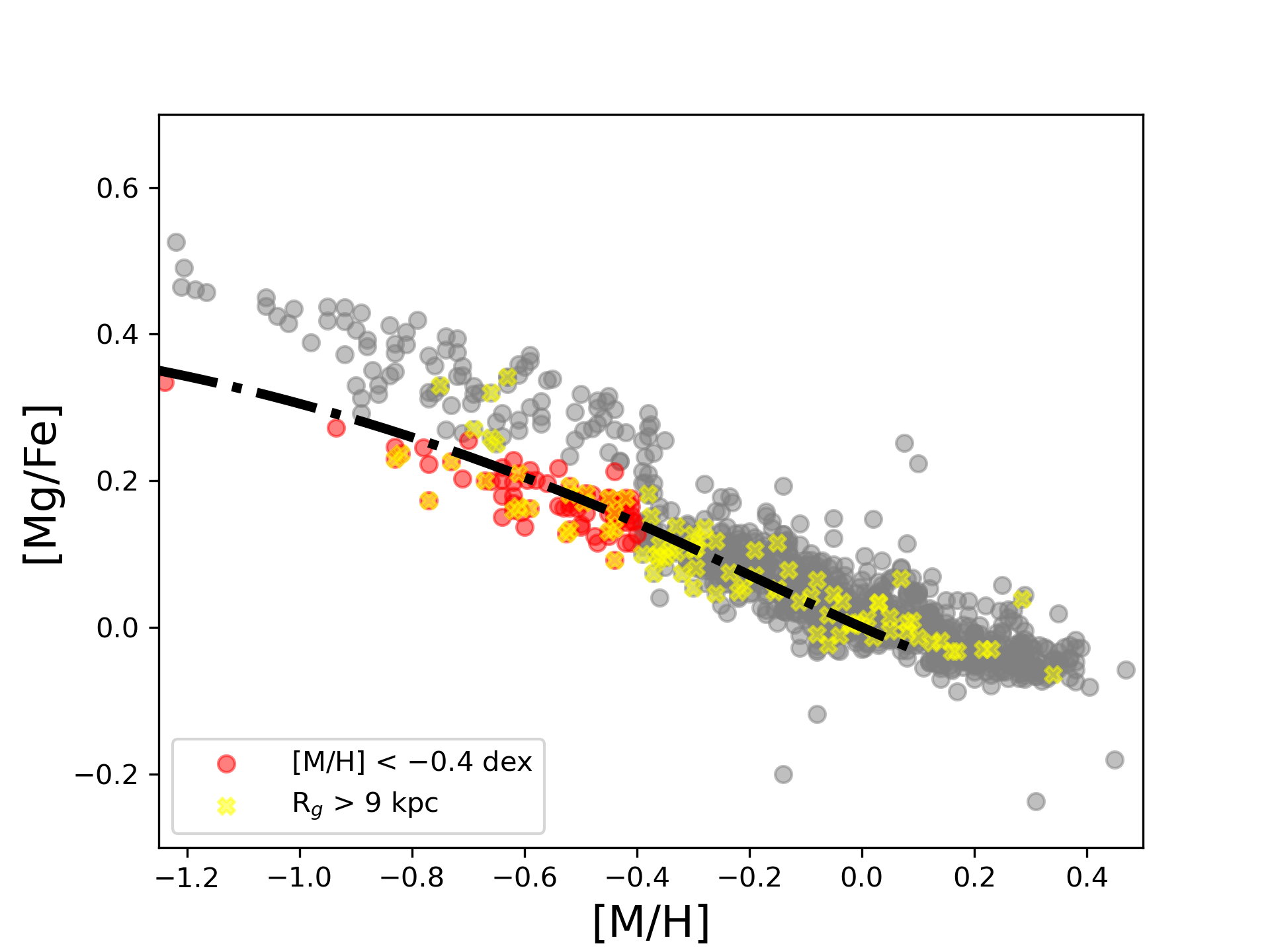}
    \caption{[Mg/Fe] versus [M/H] abundance diagram for the delayed two-infall (upper panel) and the one-infall (lower panel) models for outer Galactocentric radii (${\rm R}=$12 kpc, dash-dotted lines). The stellar yields are from \citet{Francois04}. The red points are the observed low-$\alpha$, low metallicity stars ([M/H]<-0.4 dex), whereas the yellow crosses indicate the stars with guiding radius (${\rm R}_g$) larger than 9 kpc.}
    \label{fig:outer_radii}
\end{figure}

In Fig. \ref{fig:outer_radii}, we show the results of two-infall and one-infall models for outer (${\rm R}>$9 kpc) Galactocentric radii.\\
As can be seen in the upper panel, the standard two-infall model for outer Galactic regions (thick dash-dotted line) enables us to reproduce only part of the low-$\alpha$, low-metallicity data (red points). 
To recover a good agreement between the two-infall model and the bulk of low-$\alpha$, metal-poor stars, we need to adopt a larger ratio between the low-$\alpha$ present-day surface mass density ($\Sigma_{low}$) and the high-$\alpha$ one ($\Sigma_{high}$). 
Adopting a larger ratio between surface mass densities ensures that lower metallicities are reached during the second infall episode. This is because of the larger proportion of pristine gas, which favours the presence of a more prominent loop feature (see \citealt{Spitoni19}). For this reason, we run a model where we increase $\Sigma_{low}/\Sigma_{high}$ by a factor 3 (thin dash-dotted line), leaving unchanged the other prescriptions. 
Despite the $\lq$ad hoc' prescription, this model is physically motivated by the observations coming from large surveys at large galactocentric radii (e.g. \citealt{Queiroz20}).\\
In the lower panel instead, we see that the one-infall model for outer radii naturally reproduces the tail of the low-$\alpha$ data distribution. To test even more the reliability of this solution, in Fig. \ref{fig:outer_radii} we highlight the stars from \citetalias{Santos20,Santos21} with guiding radius ${\rm R}_g$ larger than 9 kpc (yellow crosses). These stars are found in the solar neighbourhood thanks to their quite eccentric orbits, and thus are ideal candidates for being blurred stars from the outer regions.
As can be seen in Fig. \ref{fig:outer_radii} lower panel, the one-infall model for outer Galactocentric radii reproduces the abundance trend for most of the data with large ${\rm R}_g$. 

\begin{figure}
    \centering
    \includegraphics[width=1.\columnwidth]{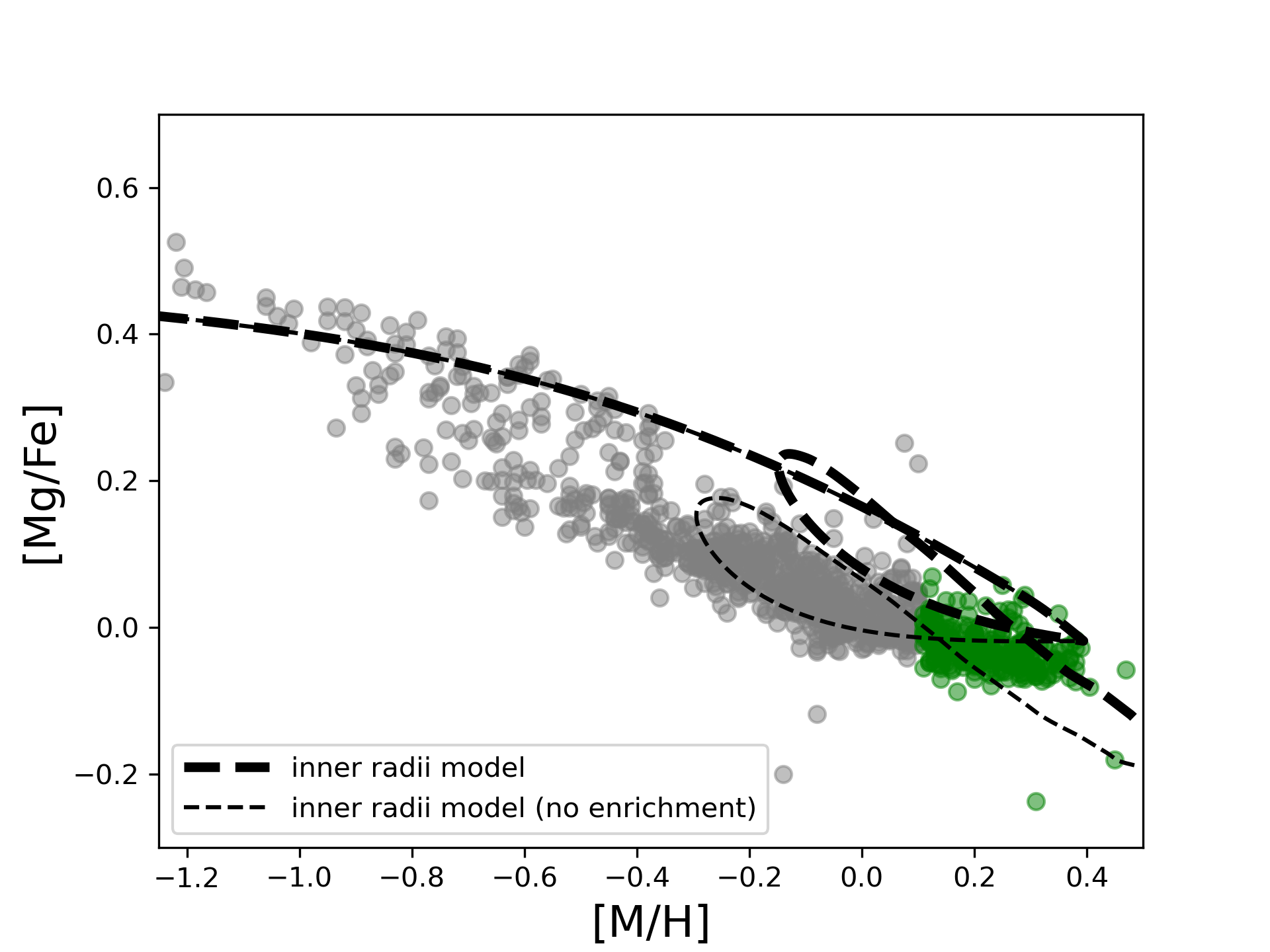}
    \includegraphics[width=1.\columnwidth]{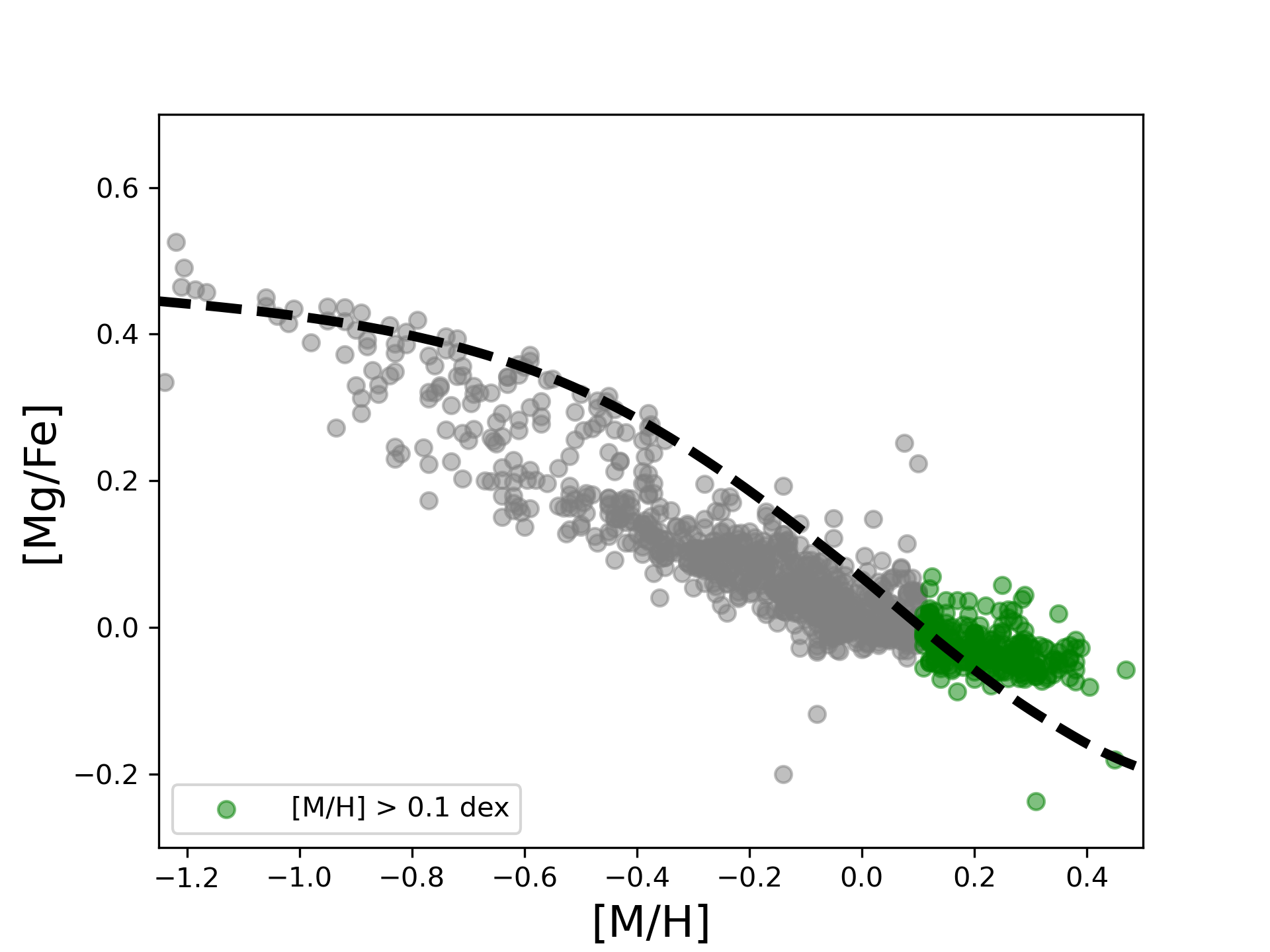}
    \caption{Same of Fig. \ref{fig:outer_radii}, but for models for inner Galactocentric radii (${\rm R}=$4 kpc, dashed lines). The stellar yields are from \citet{Francois04}. Green points indicate the observed SMR stars ([M/H]>0.1 dex).}
    \label{fig:inner_radii}
\end{figure}

In Fig. \ref{fig:inner_radii},  we do the same as in Fig. \ref{fig:outer_radii} but for the two-infall and one-infall models for inner (${\rm R}<$ 7 kpc) regions of the disc. In order to highlight the region where the model for the solar neighbourhood struggles to reproduce the data, we plot in green the observed SMR stars.\\
In Fig. \ref{fig:inner_radii} upper panel, 
we show the chemical tracks obtained with the two-infall model for inner radii. In particular, we display a model without second infall pre-enrichment (thin dashed line) and a model with the second infall enriched at [Fe/H]=-0.75 dex level (thick dashed line). 
It is worth noting that metal-enriched gas accretion was suggested by \citet{Palla20} and \citet{Spitoni21} to explain APOGEE data (\citealt{Hayden15,Ahumada19}) for inner regions of the MW disc. They argue that this infall enrichment could originate from the formation of the thick disc, Galactic halo, Galactic bar, or previous merger events, which then gets mixed with a larger amount of infalling primordial gas, as also suggested by several cosmological simulations (\citealt{Renaud20,Khoperskov21}).\\
The chemical track of the model with enrichment in the second gas accretion episode better reproduces the bulk of the SMR stars, as the enrichment in the second infall enables a larger rise in the [$\alpha$/Fe] (see also \citealt{Palla20}). In fact, without any pre-enrichment, the model tends to predict a bit too low [$\alpha$/Fe] for a given metallicity.\\
For what concerns Fig. \ref{fig:inner_radii} lower panel, we note that the one-infall model has similar problems to the two-infall model without infall pre-enrichment. The predicted [Mg/Fe] for [M/H]$\gtrsim$0.2 dex are generally lower than what observed for SMR stars.\\

\subsubsection*{Considering stellar radial migration}

\begin{figure*}
    \centering
    \includegraphics[width=0.9\columnwidth]{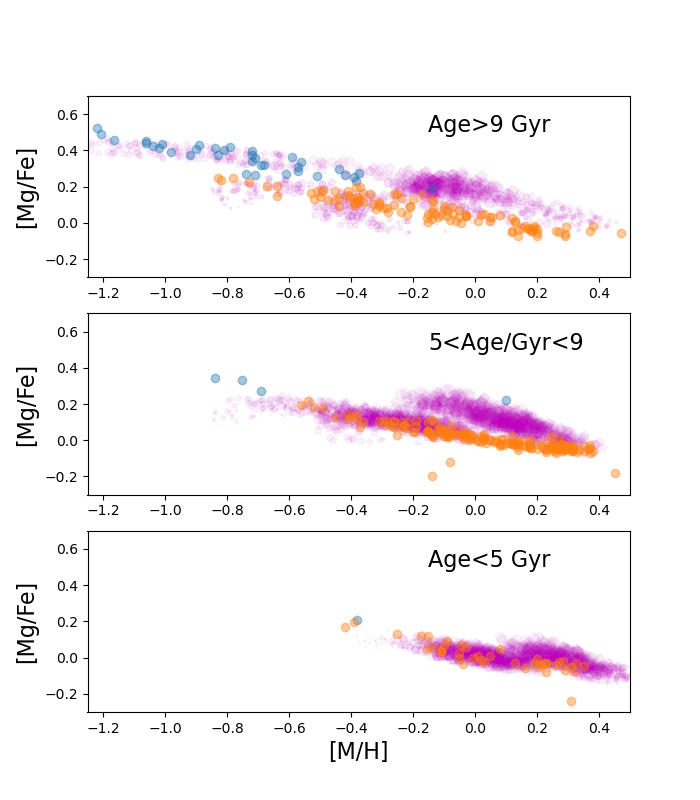}
    \includegraphics[width=0.9\columnwidth]{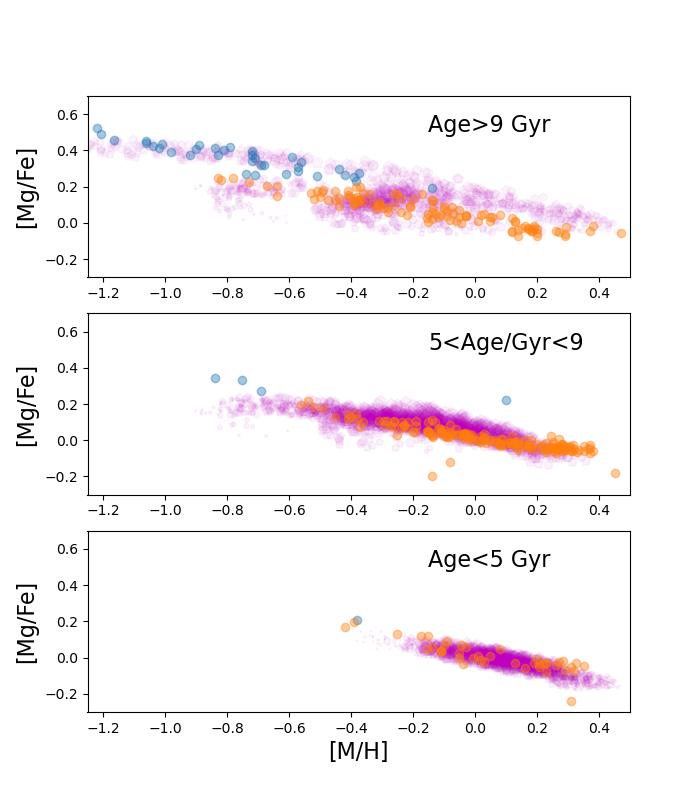}
    \caption{Synthetic model results for [Mg/Fe] versus [M/H] distributions including stellar migration produced by the delayed two-infall model with second infall enrichment for inner radii (left panels) and without second infall enrichment (right panels).  Radial migration prescriptions from \citet{Frankel18}. The stellar yields are from \citet{Francois04}. Purple filled pentagons are the mock data from the synthetic model. Mock and observational data for the MSTO subsample are plotted as in Fig. \ref{fig:age_MH_2inf}, with density normalisation done independently for each panel. Upper panels: stars older than 9 Gyr. Middle panels: stars with age between 9 and 5 Gyr. Bottom panels: stars younger than 5 Gyr.}
    \label{fig:migration_twoinfall}
\end{figure*}

\begin{figure}
    \centering
    \includegraphics[width=0.9\columnwidth]{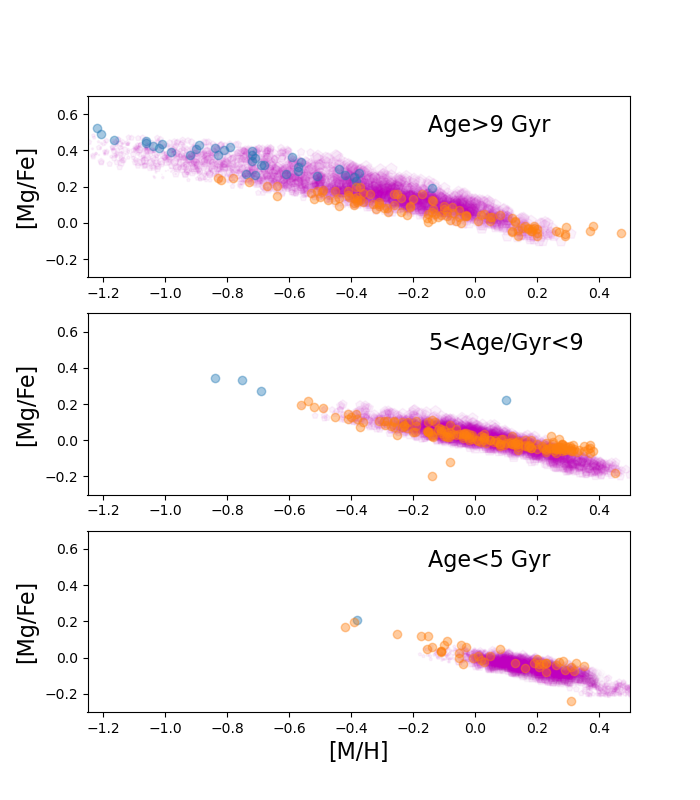}
    \caption{Same of Fig. \ref{fig:migration_twoinfall}, but for the one-infall model. The stellar yields are from \citet{Francois04}. Radial migration prescriptions are from \citet{Frankel18}.}
    \label{fig:migration_parallel}
\end{figure}

To make a step forward in our analysis, the chemical tracks described in this Section are implemented with detailed prescription for stellar radial migration (\citetalias{Spitoni15,Frankel18}). 
It is worth noting that similar results are found adopting these two prescriptions for both the two-infall (Fig. \ref{fig:migration_twoinfall}) and the one-infall (Fig. \ref{fig:migration_parallel}) scenarios. For this reason, we decide to show only the results obtained with the most recent  \citetalias{Frankel18} prescriptions.\\

In Fig. \ref{fig:migration_twoinfall} and \ref{fig:migration_parallel} we show evolution of [Mg/Fe] vs. [M/H] considering radial migration effect at different age bins. This plot can deliver fundamental information about stars in our sample. In fact, we need to fit both abundance and age data simultaneously by adopting the MSTO subsample by \citetalias{Santos21}.
To obtain a fair comparison with the data, the stellar selection function and observational errors are taken into account in the models, as in \ref{sss:ages}.

In Fig. \ref{fig:migration_twoinfall} left panels, we show the results for the two-infall model adopting a second infall pre-enrichement for R<7 kpc. It is evident that this latter assumption causes the presence of a large number of high-$\alpha$ stars at high metallicity, which are not observed in the MSTO subsample nor in the total sample by \citetalias{Santos20,Santos21}. 
The problem does not hold anymore if we do not assume an enrichment in the second gas accretion (Fig. \ref{fig:migration_twoinfall} right panels), with all the predicted stars lying in the low-$\alpha$ sequence in the two most recent age bins. However, it is worth noting that in the oldest age bin, this model has problems in predicting the low-$\alpha$ SMR stars, even considering the observational uncertainties.\\
On the other hand, the model is able to reproduce the low-metallicity tail of low-$\alpha$ stars at different age bins. This is possible thanks to the increased ratio between low-$\alpha$/high-$\alpha$ surface mass density, which favours the presence of a more prominent loop feature at the start of the second infall episode and so the presence of more metal poor stars.

For what concerns the one-infall model in Fig. \ref{fig:migration_parallel}, we see that the mock data reproduce the bulk of MSTO stars in all age bins. However, we also note that, although we take into account stellar migration from inner and outer regions, the model still struggles to reproduce the low metallicity tail of the low-$\alpha$ sequence, especially in the young and intermediate-age bins.

\begin{figure}
    \centering
    \includegraphics[width=0.9\columnwidth]{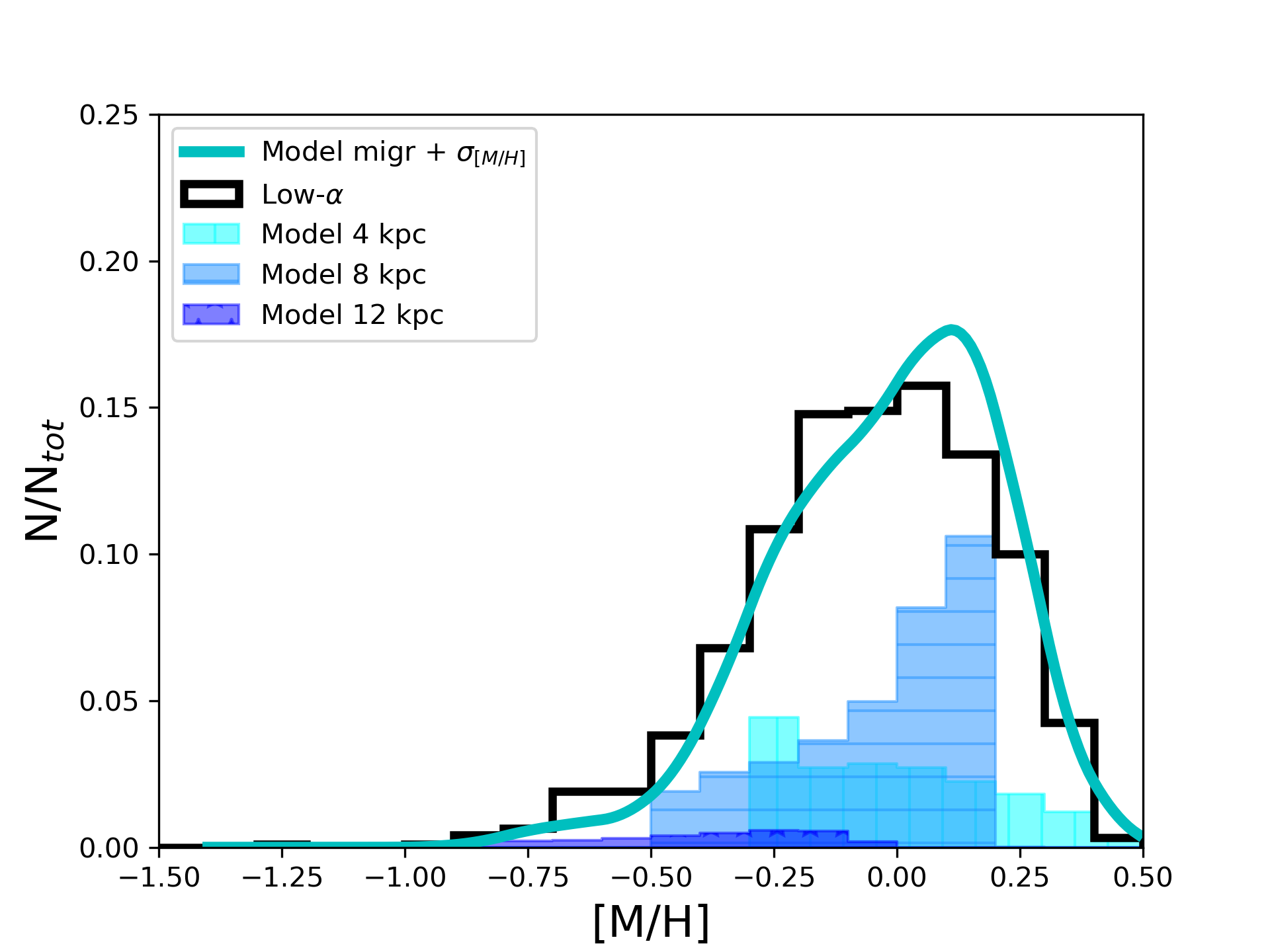}
    \includegraphics[width=0.9\columnwidth]{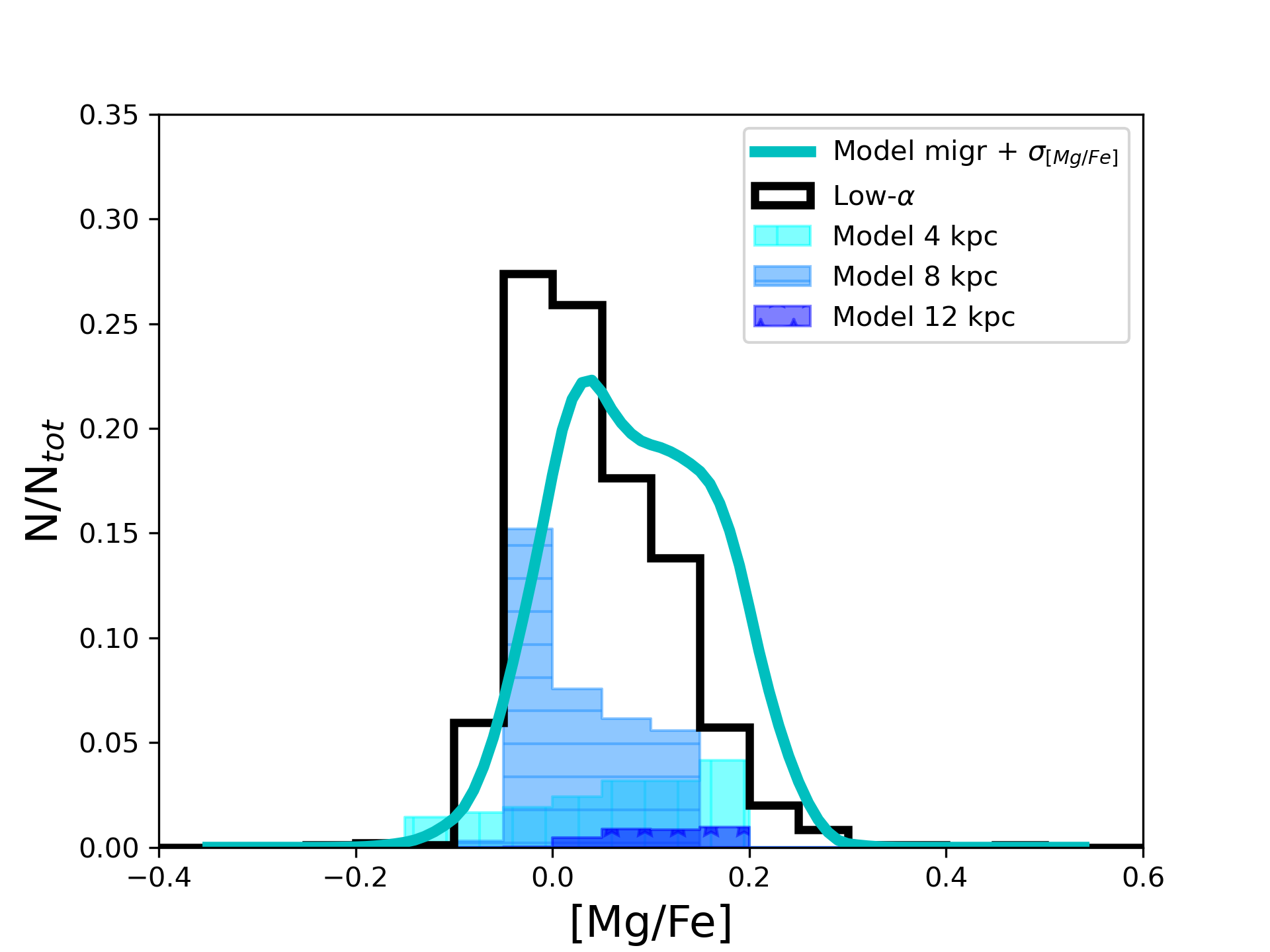}
    \caption{[M/H] and [Mg/Fe] distribution functions observed for low-$\alpha$ stars (black histogram) compared with the theoretical distribution ones for the two-infall model including migration prescriptions from \citet{Frankel18} (cyan line) and corrected for the selection function. The theoretical distribution functions are  convoluted with Gaussians with $\sigma_{[M/H]}$ and $\sigma_{[Mg/Fe]}$ (see \ref{sss:ages}).
    The cyan, light blue and blue shaded regions indicate the contribution from stars born around 4 kpc, 8 kpc and 12 kpc, respectively.}
    \label{fig:MDF_twoinfall}
\end{figure}

\begin{figure}
    \centering
    \includegraphics[width=.9\columnwidth]{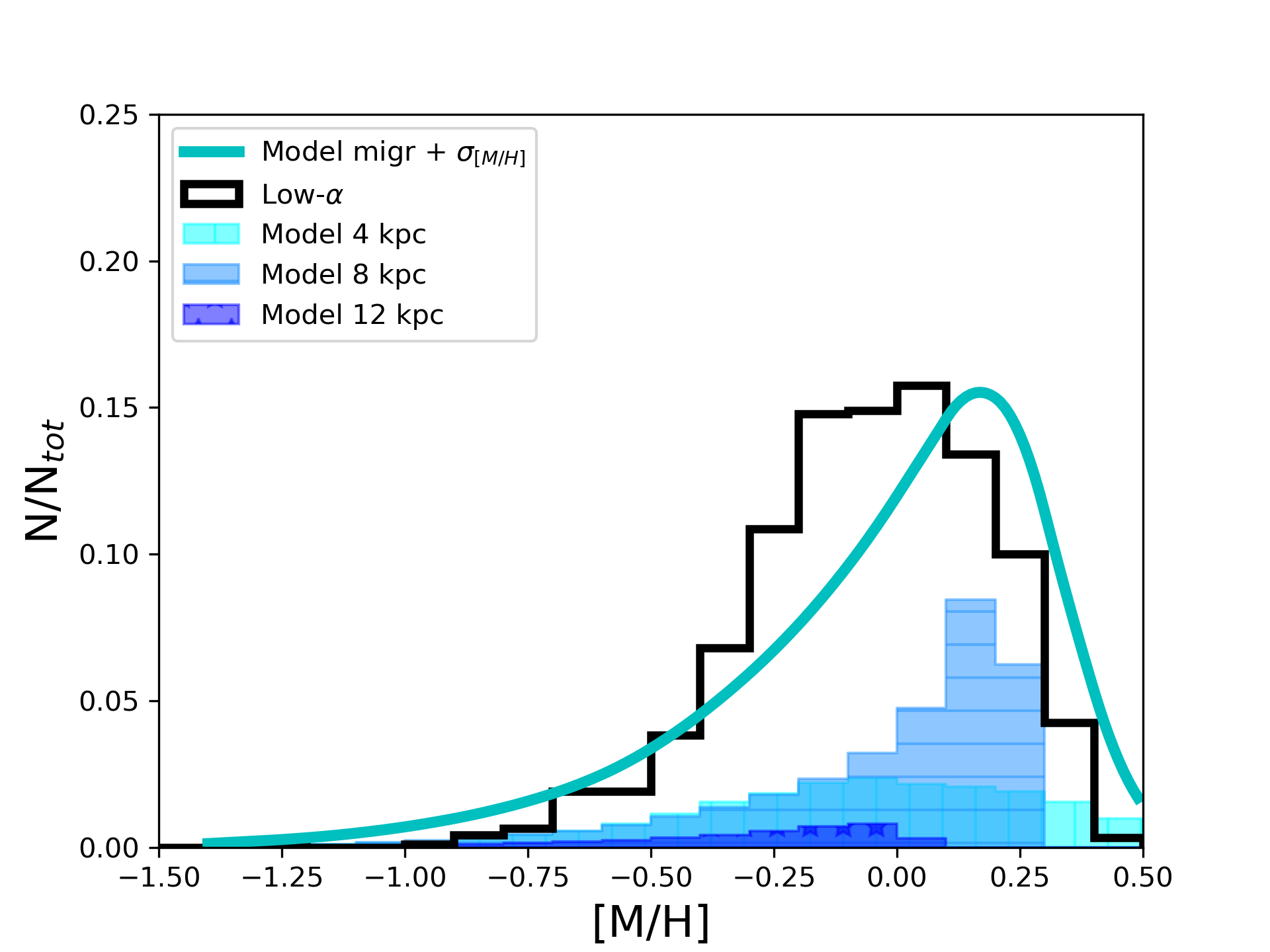}
    \includegraphics[width=.9\columnwidth]{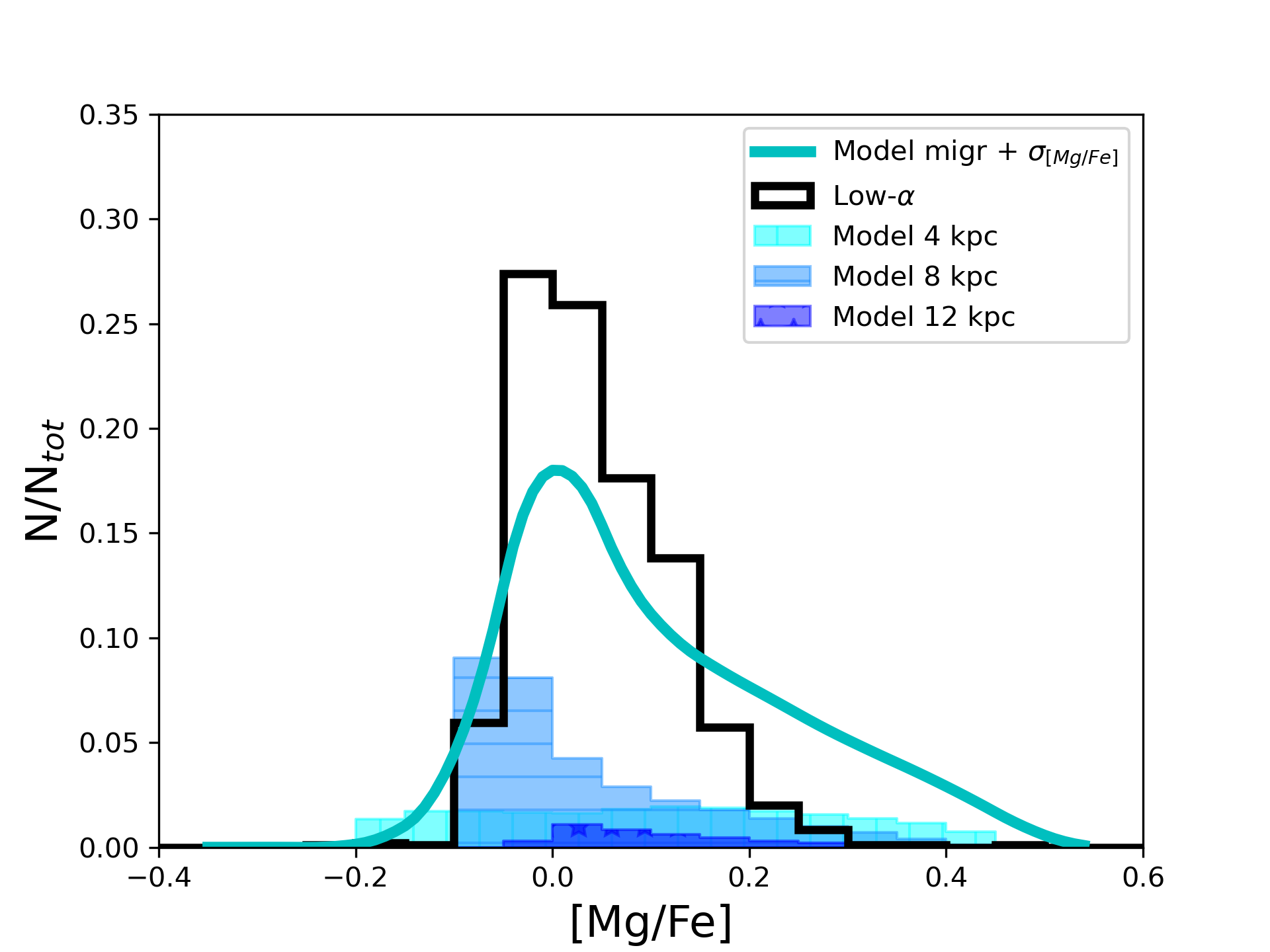}
    \caption{Same of Fig. \ref{fig:MDF_twoinfall}, but for the one-infall model. The stellar yields are from \citet{Francois04}. Radial migration prescriptions are from \citet{Frankel18}.}
    \label{fig:MDF_parallel}
\end{figure}

The [M/H] and [Mg/Fe] distribution functions for the two-infall and one-infall models  
are shown in Figs. \ref{fig:MDF_twoinfall} and \ref{fig:MDF_parallel}, respectively. In addition to the global distribution function (that take into account the observational uncertainties), in each panel we show the contribution from the stars born at different Galactocentric radii bins.

Starting from the two-infall model (assuming no enrichment in the second infall for the inner radii), we see that the "global" 
[M/H] distribution (Fig. \ref{fig:MDF_twoinfall} top panel) is well reproduced: the peak in the model distribution coincides with that from the data and no lacks of stars at high [M/H] are observed. However, a partial deficiency of the low-metallicity tail contribution is still seen for the two-infall model. This can be solved by: i) assuming more stars migrated from outer radii, or ii) adopting a larger $\Sigma_{low}$ in the outer regions of the disc, which corresponds to a flatter density profile for the low-$\alpha$ population in the outer radii. This last solution can be also seen in the context of the larger $\Sigma_{low}/\Sigma_{high}$ ratio needed to reproduce the low-$\alpha$, low metallicity data, providing an alternative explanation to that of a suppressed high-$\alpha$ population density in the outermost disc.\\
Also the [Mg/Fe] distribution function (Fig. \ref{fig:MDF_twoinfall} bottom panel) is well reproduced by the two-infall model, with the peak of the model distribution that coincides with that of the data. In this case we note a slight overabundance of low Mg stars relative to the sample of \citetalias{Santos20,Santos21}. This is due to the contribution of the stars from the innermost radii, as we can see from Fig. \ref{fig:MDF_twoinfall} bottom panel (see also Fig. \ref{fig:inner_radii}). This happens despite of the fact that we adopt the \citetalias{Francois04} yield set, that was the only one that does not predict a consistent [Mg/Fe] underestimation at high metallicity (see \ref{ss:yield_result} and \ref{ss:scenario_result}).

For the one-infall model in Fig. \ref{fig:MDF_parallel}, we see that the scenario reproduces quite well the observed [M/H] distribution (Fig. \ref{fig:MDF_parallel} top panel). However, the predicted peak is shifted towards higher metallicities, and the situation does not change even when we consider less efficient radial migration. In fact, the 8 kpc bin has a peak in the distribution function at values similar to those shown in Fig. \ref{fig:MDF_parallel}.\\
However, larger problems are seen in the [Mg/Fe] distribution function (Fig. \ref{fig:MDF_parallel} bottom panel). In particular, the one-infall model predicts too many stars with high Mg. This is the consequence of having a model with primordial gas accretion.

Therefore, we can say that the two-infall model is the best to reproduce the observed low-$\alpha$ stellar distribution in \citetalias{Santos20,Santos21}.
Moreover, the predicted [Mg/Fe] distribution function by the one-infall scenario points toward an overestimation of the high [Mg/Fe] stars even including the high-$\alpha$ data of \citetalias{Santos20,Santos21} (which should be already explained by the other model sequence).\\

\begin{figure}
    \centering
    \includegraphics[width=.9\columnwidth]{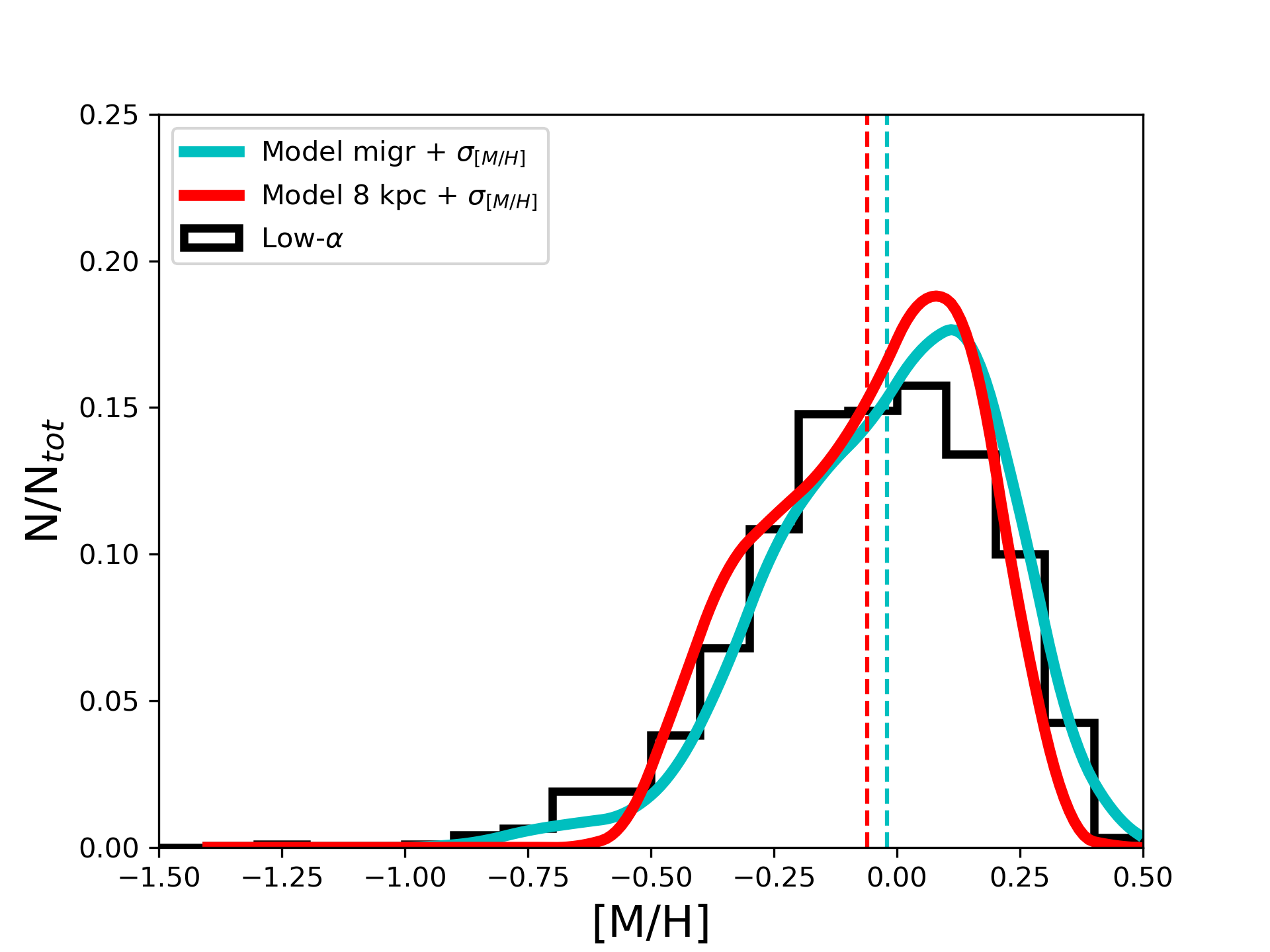}
    \includegraphics[width=.9\columnwidth]{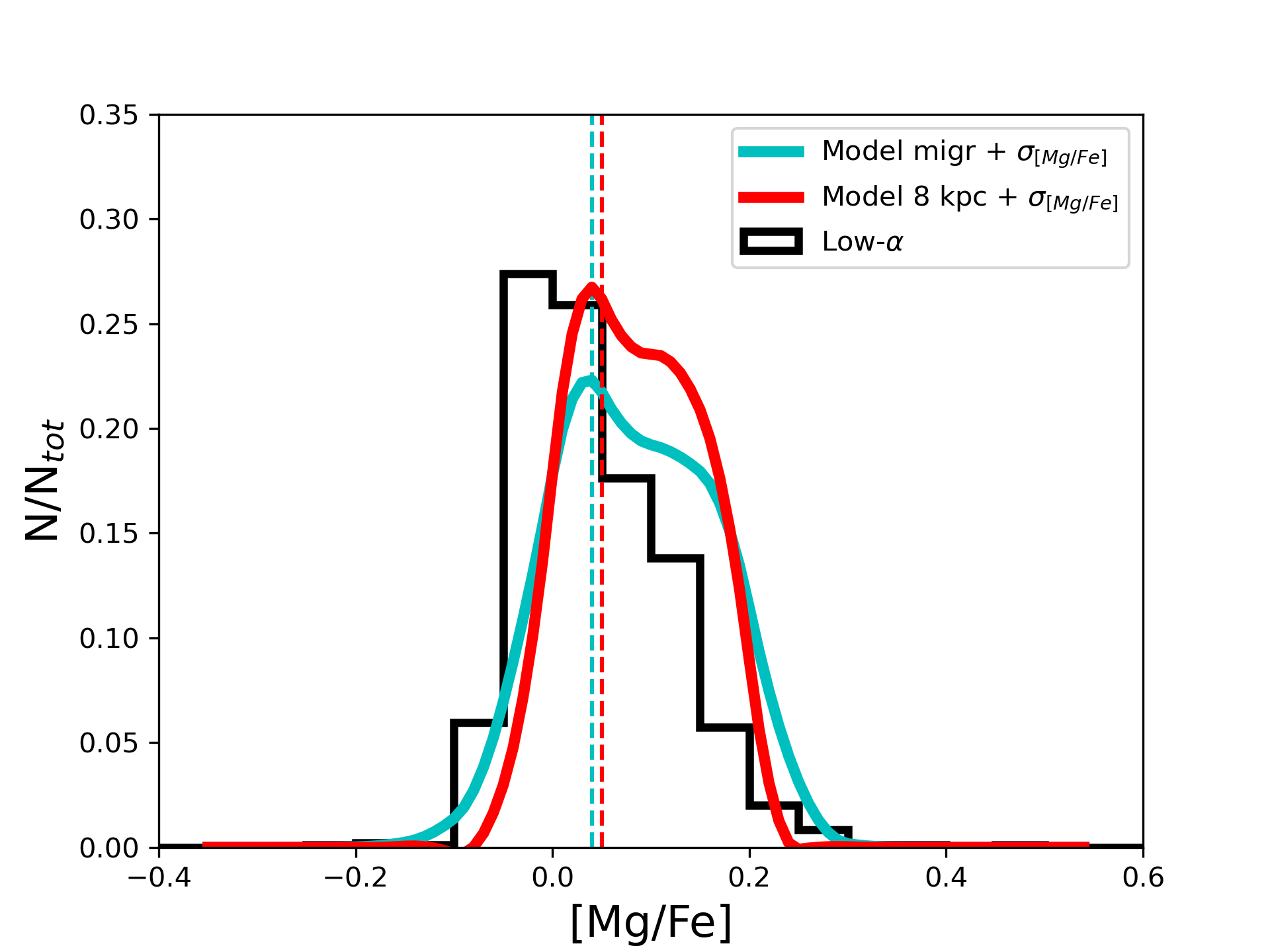}
    \caption{[M/H] and [Mg/Fe] distribution functions observed for low-$\alpha$ stars (black histogram) compared with the theoretical ones for the two-infall model including migration prescriptions from \citet{Frankel18} (cyan line) and no migration (red lines). Vertical dashed lines are the median [X/Y] ratios for models with (cyan) and without migration (red).}
    \label{fig:MDF_compare}
\end{figure}

In Fig. \ref{fig:MDF_compare} we show the impact of radial migration in shaping the distribution functions. In particular, we show the results obtained with the two-infall model; however, we can make similar considerations if we plot the results for the one-infall model.

What we note is that, overall, the predicted stellar distributions are not substantially affected by migration. In particular, the positions of the peaks are left roughly unchanged and the same happens if we look at the predicted median [M/H] and [Mg/Fe] ratios (see vertical dashed lines). \\
However, radial migration is needed in order to improve the agreement between model predictions and observations. In particular, this is necessary to explain the low- and high-metallicity tails of the [M/H] distribution, even if it is worth reminding that the high [M/H] may be biased towards higher values due to sample selection effects out of the usual $\log\, g$, $T_{eff}$ cuts (see \citetalias{Santos21}).

Nevertheless, from Fig. \ref{fig:MDF_compare} it is evident that the stellar distribution functions cannot be used to extract information on the extent and significance of stellar migration.
For this reason, it is very hard to quantify the impact of stellar radial migration in the sample by \citetalias{Santos20,Santos21}.\\
The need for models describing the evolution at different Galactocentric radii to explain the trends of the data confirms that radial migration actually plays a role in shaping the abundance observed in the solar neighbourhood, but the information we can extract from the previous Figures cannot say much about the fractional contribution from stars born outside the solar vicinity.
In fact, the migration prescriptions adopted in this Subsection predict between 50\% to 30\% of stars observed born in the solar neighbourhood, but the results about the [Mg/Fe] vs. [M/H] evolution and the distribution functions are rather similar.

\section{Discussion and conclusions}
\label{s:conclusions}

In this paper, we compare the new [Mg/Fe] abundance measurements by \citet{Santos20,Santos21} with detailed chemical evolution models for the solar vicinity, exploring different stellar yields (\citealt{Francois04,Koba06,Limongi18}) and chemical evolution scenarios (delayed two-infall, e.g. \citealt{Palla20}, and parallel, e.g. \citealt{Grisoni17}). In the light of the obtained results, we also consider the possible impact of stellar radial migration on the adopted sample by implementing migration prescriptions from the literature (\citealt{Spitoni15,Frankel18}).\\
In \citet{Santos20} it was performed an optimised normalization procedure for the high-resolution stellar spectra obtained by HARPS in the context of the AMBRE project (\citealt{deLaverny13}). The new procedure allows a significant improvement in the abundance determination precision, allowing to see a decreasing trend in [Mg/Fe] with metallicity at high metallicity, at variance with previous works (e.g. \citealt{Adibekyan12,Hayden17}). In this way, it is possible to test whether the discrepancy between the observed flat trend in the metal-rich disc and the steeper slope predicted by the models can be solved by these abundance measurements. The value of this data set was even enhanced in \citet{Santos21}, in which reliable stellar age estimates were presented for a subsample of MSTO stars. All the \citet{Santos20,Santos21} data are cross-matched with the $Gaia$ DR2 catalogue (\citealt{Gaia18}) in order to extract the main kinematical and dynamical stellar parameters.\\
The comparison between the wealth of chemical, age and orbital information has allowed us to place constraints not only on the Mg stellar production by stellar sources and on the evolutionary scenarios describing the solar vicinity, but also to describe the influence of stellar migration within few hundreds of parsecs from the Sun. In this way, we also test which histories of chemical evolution should have experienced the inner and outer Galactic regions.\\

The main results of this work can be summarized as follows:
\begin{enumerate}
    \item the adoption of different sets of massive star stellar yields have a crucial impact on the predicted chemical evolution. To recover a good agreement with \citet{Santos20,Santos21} data, we have still to rely on the semi-empirical yields by \citet{Francois04}. Acceptable results are still obtained using the \citet{Koba06} yield set, even though a steeper [Mg/Fe] trend at high metallicity is predicted relative to the new abundance data, claiming either a revision of CC-SN yields or further biases in the abundance determination (despite of the optimised derivation shown in \citealt{Santos20}).
    The various nucleosynthesis results proposed \citet{Limongi18} have instead severe problems in recovering the observed [Mg/Fe] both at low and high metallicity, even adopting the most Mg rich yield sets.\\
    The robustness of these results relative to model parameter variations suggests to take with caution the choice of the yield set when dealing with Mg and that further work must be done in the context of the stellar nucleosynthesis for this element;
    
    \item both the delayed two-infall and the parallel scenarios of chemical evolution satisfactorily reproduce the bulk of \citet{Santos20,Santos21} data in the [Mg/Fe] vs. [M/H] plane for both the high-$\alpha$ and the low-$\alpha$ sequences. However, they both encounter problems in explaining the metal rich tail of the data ([M/H]$\gtrsim$0.1-0.2 dex) as well as the the most metal poor stars of the low-$\alpha$ sequence.\\
    In order to test these results on firmer bases, we compare our models to the \citet{Santos21} MSTO subsample with stellar ages. Even applying an $\lq$a posteriori' error on the chemical evolution tracks (see Section \ref{ss:scenario_result}), the models struggle to reproduce the metal-rich and metal-poor tails of the low-$\alpha$ sequence, indicating that these stars have experienced radial migration from their original birthplace;
    
    \item we test the origin of the migrated stars in the light of the one-infall and delayed two-infall scenarios by taking advantage of the prescriptions by \citet{Grisoni18} and \citet{Palla20} for inner and outer disc chemical tracks. \\
    We find that low-$\alpha$, low metallicity stars are likely to be outer disc candidates. These stars are well explained by a two-infall model with prescription suited for the outer disc (e.g. larger infall timescale due to inside-out, reduced star formation efficiency) and with negligible high-$\alpha$ population surface mass density at large Galactocentric distances. In particular, this latter feature can be seen in the context of the lack outer disc stellar data by large surveys (e.g. \citealt{Queiroz20}). However, we cannot exclude larger low-$\alpha$ population surface densities than those predicted for the outer disc to be the cause of this behaviour. \\
    Concerning the high metallicity end of the low-$\alpha$ sequence, we can explain these stars in terms of stars migrated from the inner disc. At variance with previous works (e.g. \citealt{Palla20}), we exclude that inner disc regions are formed by an enriched gas accretion episode. However, it is worth reminding that different datasets as well as stellar yields prescriptions were adopted in these previous papers.\\
    We also find that the one-infall model including migration prescriptions has problems in reproducing the shape of the stellar distribution functions and in particular that of [Mg/Fe]; 

    \item despite the evidence found for stellar radial migration, it is difficult to give an estimation of the faction of stars migrated to the solar vicinity from other part of the MW disc to explain the sample of \citet{Santos20,Santos21}. Different prescriptions for radial migration have small effect on the overall distribution functions, with different fractions of stars from inner and outer Galactocentric rings giving similar overall results. Therefore, these are not good tools to impose constraints on migration.\\
    This leaves the door open for different conclusions about the contribution of migration on the MW disc (e.g. \citealt{Vincenzo20,Sharma20}).
    However, our results highlight that peculiar histories of star formation, such as that of the two-infall model are needed to explain the distribution of stars in the Galaxy, as also suggested by several works (e.g. \citealt{Mackereth18,Khoperskov21}, see also \citealt{Vincenzo21}).
    
\end{enumerate}

\begin{acknowledgements}
     The authors thank the referee for the careful reading of the manuscript and the comments that significantly improve the quality of the paper.\\
     M.P. acknowledges the computing centre of CINECA and INAF under the coordination of the "accordo quadro MoU per lo svolgimento di attività congiunta di ricerca nuove frontiere in astrofsica: HPC e data exploration di nuova generazione" for the availability of computing resources. \\
     P.S.P. would like to thank the Centre National de Recherche Scientifique (CNRS) for the financial support, and partial support from the Université Côte d'Azur (UCA). P.S.P also acknowledges financial support by the Spanish Ministry of Science and Innovation through the research grant PID2019-107427GB-C31. This work has made use of data from the European Space Agency (ESA) mission {\it Gaia} (\url{https://www.cosmos.esa.int/gaia}), processed by the {\it Gaia} Data Processing and Analysis Consortium (DPAC, \url{https://www.cosmos.esa.int/web/gaia/dpac/consortium}). Funding for the DPAC has been provided by national institutions, in particular the institutions participating in the {\it Gaia} Multilateral Agreement.
\end{acknowledgements}


\bibliographystyle{aa}
\bibliography{MgFe_paper}

\end{document}